\documentclass{IEEEtran4PSCC}
\ifCLASSINFOpdf
   \usepackage[pdftex]{graphicx}
\usepackage{enumerate}
\else
   \usepackage[dvips]{graphicx}
\fi
\usepackage[font=small]{caption}

\usepackage[cmex10]{amsmath}
\usepackage[noadjust]{cite}
\usepackage{enumerate}
\usepackage{color}
\usepackage{xcolor}
\usepackage{amsmath, amssymb}
\usepackage{enumerate}
\DeclareMathAlphabet\mathbfcal{OMS}{cmsy}{b}{n}

\usepackage{array}
\usepackage{tabularx}
\usepackage{optidef} 
\usepackage{booktabs}
\usepackage{mathtools}

\usepackage{url}
\usepackage{amsthm}
\usepackage{optidef}
\usepackage{mathtools}
\usepackage{subcaption}
\newtheorem{remark}{Remark}

\usepackage{makecell}
\usepackage{hyperref}
\usepackage{fancyhdr}
\usepackage{atbegshi}
\usepackage{tikz}
\usepackage{everypage}
\usepackage[ruled,vlined,linesnumbered]{algorithm2e}
\DontPrintSemicolon
\SetKwInput{KwInput}{Input}
\SetKwInput{KwOutput}{Output}
\SetKwComment{Comment}{$\triangleright$ }{}

\renewcommand{\baselinestretch}{.99}
\makeatletter
\let\old@ps@headings\ps@headings
\let\old@ps@IEEEtitlepagestyle\ps@IEEEtitlepagestyle
\def\psccfooter#1{%
    \def\ps@headings{%
        \old@ps@headings%
        \def\@oddfoot{\strut\hfill#1\hfill\strut}%
        \def\@evenfoot{\strut\hfill#1\hfill\strut}%
    }%
    \def\ps@IEEEtitlepagestyle{%
        \old@ps@IEEEtitlepagestyle%
        \def\@oddfoot{\strut\hfill#1\hfill\strut}%
        \def\@evenfoot{\strut\hfill#1\hfill\strut}%
    }%
    \ps@headings%
}
\makeatother

\psccfooter{%
        \parbox{\textwidth}{\hrulefill \\ \small{24th Power Systems Computation Conference} \hfill \begin{minipage}{0.2\textwidth}\centering \vspace*{4pt} \includegraphics[scale=0.06]{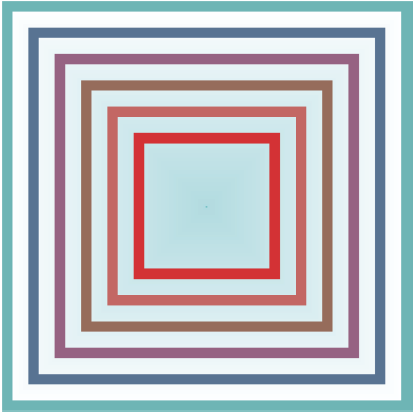}\\\small{PSCC 2026} \end{minipage} \hfill \small{Limassol, Cyprus --- June 8 -- June 12, 2026}}%
}
\usepackage{enumerate}
\usepackage{float}
\usepackage{array}
\usepackage{multirow}
\usepackage{adjustbox}
\DeclareMathAlphabet\mathbfcal{OMS}{cmsy}{b}{n}

\AddEverypageHook{%
\begin{tikzpicture}[remember picture,overlay]
  \node [rotate=-90, text=gray!80, font=\small\sffamily] 
    at ([xshift=-0.5in]current page.east) 
    {Accepted for presentation at the 24th Power Systems Computation Conference (PSCC 2026), Limassol, Cyprus};
\end{tikzpicture}}

\begin{document}
\title{AC Dynamics-aware Trajectory Optimization with Binary Enforcement for Adaptive UFLS Design}
\author{
\IEEEauthorblockN{Muhammad Hamza Ali, Amritanshu Pandey}
\IEEEauthorblockA{Department of Electrical and Biomedical Engineering \\
University of Vermont\\
Burlington, Vermont}
}
\maketitle
\begin{abstract}
The high penetration of distributed energy resources, resulting in backfeed of power at the transmission and distribution interface, is causing conventional underfrequency load shedding (UFLS) schemes to become nonconforming.
Adaptive schemes that update UFLS relay settings recursively in time offer a solution, but existing adaptive techniques that obtain UFLS relay settings with linearized or reduced-order model formulations fail to capture AC nonlinear network behavior.
In practice, this will result in relays unable to restore system frequency during adverse disturbances.
We formulate an adaptive UFLS problem as a trajectory optimization and include the full AC nonlinear network dynamics to ensure AC feasibility and time-coordinated control actions.
We include binary decisions to model relay switching action and time-delayed multi-stage load-shedding.
However, this formulation results in an intractable MINLP problem.
To enforce model tractability, we relax these binary variables into continuous surrogates and reformulate the MINLP as a sequence of NLPs.
We solve the NLPs with a homotopy-driven method that enforces near-integer-feasible solutions.
We evaluate the framework on multiple synthetic transmission systems and demonstrate that it scales efficiently to networks exceeding 1500+ nodes with over 170k+ continuous and 73k+ binary decision variables, while successfully recovering binary-feasible solutions that arrest the frequency decline during worst-case disturbance.
\end{abstract}

\newcommand{\TO}{Trajectory optimization}
\newcommand{\UFLS}{adaptive UFLS design problem}
\newcommand{\tos}{trajectory optimization}
\section{Introduction}
\begin{figure*}
    \centering
\includegraphics[width=\textwidth,height=4.5\textwidth,keepaspectratio]{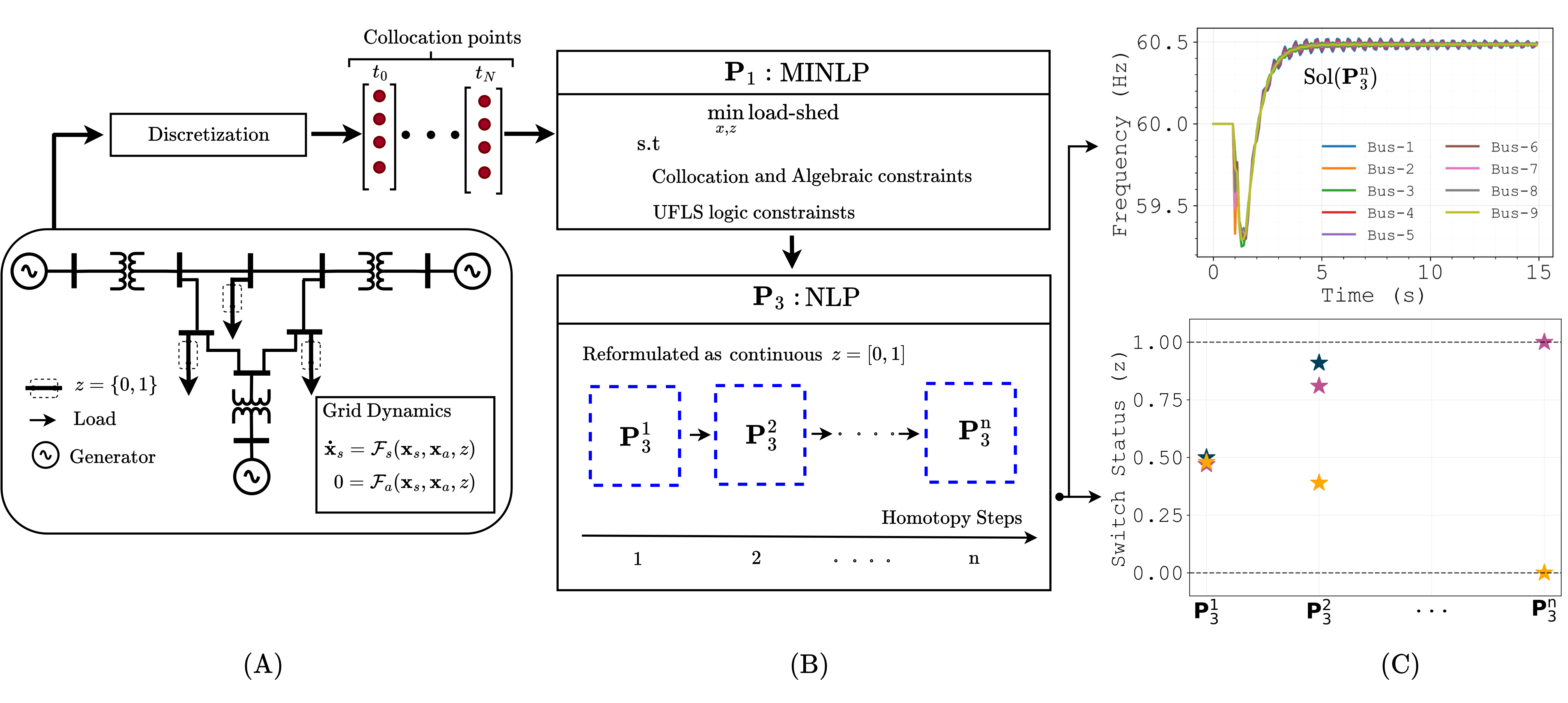}
    \caption{Illustration of our adaptive UFLS design \tos{} framework applied to the WSCC 9-bus system (A) Trapezoidal direct collocation discretizes the nonlinear AC network dynamics with binary decision variables ($z$). (B) The reformulation converts the MINLP $(\mathbf{P}_1)$ into a sequence of NLPs $(\mathbf{P}_3^1, \mathbf{P}_3^2, \ldots, \mathbf{P}_3^\text{n})$ solved through successive homotopy steps (C) A 25\% generator-loss disturbance at $t=1$s produces binary switch trajectories (moving from near 0.5 to binary \{0, 1\}) across the homotopy steps, with the final step yielding near-integer feasibility and a frequency response within acceptable bounds.}
    \label{fig:methodology}
\end{figure*}
Modern power system operations increasingly have to make decisions that consider the grid’s AC dynamics to enable more effective responses to disturbances and rapidly changing conditions \cite{soares2018survey}. 
An important subset of these decisions also requires optimal switching of grid equipment in and out of service \cite{kothari2012power} - \cite{agarwal2023continuous}. 
These decision-making optimization problems present a two-pronged challenge.
On one hand, satisfying AC grid dynamics within a decision-making optimization is challenging.
On the other hand, the complexity is further exacerbated due to the inclusion of binary variables for modeling discrete decisions, such as equipment switching and protection operations.
In this paper, we focus on one such problem, \textit{optimal} adaptive under-frequency load shedding (UFLS) design optimization, which necessitates modeling both binary decisions and AC grid dynamics.

UFLS serves as a last-resort emergency control mechanism by stabilizing the system during severe under-frequency conditions by shedding load and restoring balance between generation and demand \cite{kundur2007power}. 
UFLS schemes are broadly classified into two main categories: (i) \textit{conventional UFLS}, which sheds predefined amounts of load once certain frequency thresholds are reached \cite{alhelou2020overview}
, and (ii) \textit{adaptive UFLS}, which dynamically adjusts the load shedding settings based on the near real-time Rate of Change of Frequency (RoCoF) and/or frequency deviation from nominal \cite{1178794} - \cite{terzija2006adaptive}. 

Conventional UFLS setpoints are becoming increasingly ineffective due to changing grid conditions. Consider, for example, a sunny day in Vermont, U.S., when many of the feeders are backfeeding: the net load was roughly 100 MW, with distributed energy resources supplying nearly 80\% of that generation \cite{NREL_GridLibrary_2024}. Conventional UFLS applies fixed frequency thresholds without accounting for backfeeding. In networks with high DER penetration like Vermont, relays can trip at the backfeeding feeders and inadvertently disconnect generation instead of shedding load. This misoperation represents a major drawback of conventional UFLS. 

Adaptive UFLS schemes serve as an alternative to conventional UFLS schemes. Prior work in the literature broadly categorizes these adaptive approaches into two groups: response-based schemes, which rely on real-time system measurements \cite{golpira2022data, he2019decentralized, 7337458}, and optimization-based schemes, which use recursive in-time mathematical analysis to determine optimal load-shedding locations \cite{ceja2012under, amraee2017probabilistic, 9599468, rafinia2020stochastic}. 
The work in \cite{golpira2022data} proposes a data-driven approach that estimates frequency nadir, RoCoF, system inertia, and power imbalance from PMU measurements to guide real-time load shedding. Similarly, \cite{he2019decentralized} develops a fully decentralized COI-based UFLS scheme but neglects voltage dependency and fast transients, while \cite{7337458} introduces a predictive method that forecasts frequency violations and load-shed amounts based on PMU data. The critical drawback of response-based methods is that these methods rely heavily on COI frequency and accurate measurements, which makes them highly vulnerable to errors from noise and large frequency swings in low-inertia systems with high DER penetration \cite{alhelou2020overview}.

Several studies formulate adaptive UFLS as an explicit optimization problem, which they solve every predetermined period of time. \cite{ceja2012under,amraee2017probabilistic} apply a mixed-integer linear program (MILP) to determine optimal frequency thresholds and load-shedding decisions, and \cite{amraee2017probabilistic} extends this to a probabilistic MILP with RoCoF relays to capture distributed generation uncertainty. To address renewable variability, \cite{rafinia2020stochastic} proposes a multi-stage stochastic robust optimization model, while \cite{9599468} presents a probabilistic co-optimization of UFLS with inertia and frequency regulation services. 

More broadly, we find that both response-based \cite{golpira2022data, he2019decentralized, 7337458} and optimization-based methods \cite{ceja2012under, amraee2017probabilistic, 9599468, rafinia2020stochastic} in current literature rely on aggregated single-generator frequency models and network-agnostic assumptions to be computationally tractable. Such simplifications overlook AC power flow feasibility, nonlinear network dynamics, and the impact of voltage on load behavior, which becomes critical with voltage-dependent loads. However, \cite{9599468} ignores the time-domain nature of UFLS, which requires discrete load-shedding actions.  
This work emphasizes the need to incorporate AC nonlinear dynamics when developing adaptive UFLS settings. In particular, we consider voltage-dependent loads that significantly impact the effectiveness of load shedding and frequency recovery due to their nonlinear response to voltage fluctuations \cite{4299516} \cite{elsaadany2025ac}.
 
We formulate the adaptive UFLS design problem as a \tos{} problem and discretize the differential-algebraic equations using the trapezoidal direct collocation method \cite{kelly2017introduction}. Figure \ref{fig:methodology} illustrates our framework.
Fig.~\ref{fig:methodology}(A) shows the discretization and generation of collocation points at each time step, which we use as collocation constraints to formulate $\mathbf{P}_1$, a Mixed-Integer Nonlinear Program (MINLP Formulation) arising from the discrete switch statuses. To ensure tractability, Fig.~\ref{fig:methodology}(B) shows relaxation of $\mathbf{P}_1$ into $\mathbf{P}_3$, a nonlinear program (Regularized NLP) and application of a homotopy-based method that enforces binary decisions within the continuous framework (see Section-\ref{sec:prob_for}).  Fig.~\ref{fig:methodology}(C) shows the evolution of switching state trajectories throughout the homotopy path and the optimal frequency trajectory after solving the final sub-problem $\mathbf{P}_{3}^{\text{n}}$.

To the best of our knowledge, this is the first work to solve large-scale optimization problems constrained on full AC  power system transient models with binary switching decisions.  
The major contributions of the paper are threefold:
\begin{enumerate}
    \item \textbf{\TO{} :} The proposed framework formulates the adaptive UFLS design problem as a \tos{} that directly incorporates full AC nonlinear dynamics. Unlike existing approaches that rely on linearized approximations, our framework ensures AC feasibility throughout the optimization trajectory and removes the need for separate feasibility checks.
    \item \textbf{Enforcing binary decisions:}  We propose a novel framework that transforms binary load-shedding decisions into continuous surrogates while preserving the integer feasible structure. By embedding a homotopy-driven enforcement method within an NLP formulation, the framework actively recovers binary-feasible solutions (within 0.0001\%) from the continuous relaxation and scales to 73k+ integer decision variables. 
    \item \textbf{Multi-stage UFLS Design:} We formulate a practical multi-stage UFLS design problem that captures the sequential staged nature of load-shedding actions. We model multiple stages through a combination of binary decision variables and impose strict sequential-ordering constraints through discrete differentiation, and translate the resulting solution into effective relay settings for practical UFLS implementation.
\end{enumerate}
Our approach solves a \tos{} to determine the optimal load-shedding locations and amounts at each stage, and the framework post-processes the optimized trajectory to derive the corresponding frequency thresholds, yielding a complete and implementable set of relay settings in practice.


\section{Preliminaries: Power System Transient Simulation Modeling}\label{sec:prelim}
We formulate the \UFLS{} as a \tos{}.
We consider binary switching and AC transient stability power system physics within the \tos{}.
We briefly discuss them in the following preliminaries.
\subsection{Power System Transient Model}
\subsubsection{Synchronous Generator Model}
We model the synchronous generator index $i \in \mathcal{N}_g  = \{1,\dots,N_g\}$ with the classical second-order swing equation in \eqref{eq:syn_gen}.
The model includes a constant internal voltage $E'_i$ behind a transient reactance $X_{i}^{d'}$. 
Here $N_g$ is the total number of generators, with the generator $i$ connected to bus $j \in \mathcal{N}_b = \{1,\dots,N_b\}$ and $N_b$ is the total number of buses.  
In \eqref{eq:syn_gen}, $\delta_i^{g}$ denotes the generator $i$ internal  angle, $\delta_{j}^{b}$ is the voltage angle of bus $j$, $\Delta\omega_i^{g}$ is the per-unit rotor-speed deviation of the generator $i$ on the base $\omega_0 = 2\pi f_0$, and $V_{j}^{b}$ is the bus $j$ voltage magnitude.

\begin{subequations}\label{eq:syn_gen}
\begin{align}
  \dot{\delta}_i^{g} &= \omega_0\,\Delta \omega^g_i, \label{eq:delta_dot}\\[3pt]
  \dot{\Delta \omega}_i^{g} &= \frac{1}{2H_i^{g}}\Big( P_{i}^{m} - D_i^{g}\,\Delta \omega_i^{g} - P_{i}^{e} \Big), \label{eq:omega_dot}\\[6pt]
  P_{i}^{e} &= \frac{E'_i\,V_{j}^{b}}{X_{i}^{d'}} \,\sin(\delta_{i}^{g} - \delta_{j}^{b}), \label{eq:Pe}\\[3pt]
  Q_{i}^{e} &= \frac{E'_i\,V_{j}^{b}}{X_{i}^{d'}} \,\cos(\delta_{i}^{g} - \delta_{j}^{b}) \;-\; \frac{(V_{j}^{b})^2}{X_{i}^{d'}}. \label{eq:Qe}
\end{align}    
\end{subequations}

\noindent Also, for generator $i$, \(H_i^{g}\) is the inertia constant, \(D_i^{g}\) is the damping coefficient, \(P_{i}^{m}\) is mechanical input power (p.u.), and \(P_{i}^{e}, Q_{i}^{e}\) are electrical active/reactive powers, respectively. This model neglects stator resistance and saliency, treating \(E'_i\) as a constant.

\subsubsection{Turbine-Governor Model}
We model the governor–turbine aggregate for the generator $i \in \mathcal{N}_g$ as a first-order low-pass filter with a time constant, $T^{\mathrm{gov}}_{i}$, driven by a power setpoint,$P^{\mathrm{ref}}_{i}$, as:
\begin{subequations}\label{eq:governor_equations}
\begin{align}
  \dot{P}_{i}^{m} &= \frac{1}{T^{\mathrm{gov}}_{i}}
  \left(-P^{m}_{i} + P^{\mathrm{ref}}_{i} - \frac{\Delta\omega_i^g}{R_i}\right), \label{eq:gov_dyn}\\[3pt]
  \underline{P_{i}^{m}} &\leq P_{i}^{m} \leq \overline{P_{i}^{m}}. \label{eq:gov_limits}
\end{align}
\end{subequations}
In~(\ref{eq:gov_dyn}), $R_i$ represents the droop of the governor, and \(\overline{P}^m_i\) and \(\underline{P}^m_i\) are the upper and lower bounds on the mechanical power. 
These equations collectively model the dynamic behavior of the generator and its interaction with the turbine governor.
\subsubsection{AC power balance}
To couple the machine dynamics with the network, algebraic equations impose AC power-balance constraints at each bus $i \in \mathcal{N}_b$: 

\begin{subequations}\label{eq:ac_pf}
\begin{align}
  P_{i}^{g}-P_{i}^{d}
  &= V_i^{b} \sum_{k} V_k^{b}\!\left[\,G_{ik}\cos(\delta_{ik}^{b})+B_{ik}\sin(\delta_{ik}^{b})\right], \label{eq:pf_P_ik}\\[3pt]
  Q_{i}^{g}-Q_{i}^{d}
  &= V_i^{b} \sum_{k} V_k^{b}\!\left[\,G_{ik}\sin(\delta_{ik}^{b})-B_{ik}\cos(\delta_{ik}^{b})\right]. \label{eq:pf_Q_ik}
\end{align}
\end{subequations}

\noindent where, \(G_{ik}, B_{ik}\) is the conductance and susceptance for line between nodes $i$ and $k$ while \(\delta_{ik}^{b} = \delta_{i}^{b} -\delta_{k}^{b}\) The generator aggregates at bus \(i\) are \(P_{i}^{g}=\sum_{j\in\mathcal{N}_g} P_{j}^{e}\) and \(Q_{i}^{g}=\sum_{j\in\mathcal{N}_g} Q_{j}^{e}\).

We represent each load $i \in \mathcal{N}_{l} = \{1,\dots,N_l\}$, where $N_l$ is the total number of loads, with a ZIP model. In ZIP load model\footnote{In this work, active power load is converted to the constant current load, and reactive power load is converted to the constant impedance load.} the active and reactive powers vary with the bus-voltage magnitude \(V_i^b\) relative to a nominal value \(V_i^{b,0}\) and $a$ and $b$ represent the fraction of constant power, current, and impedance of the ZIP load model.
\(P_i^{d,0}\) and \(Q_i^{d,0}\) are the nominal active and reactive powers at \(V_i = V_i^{0}\).
\begin{subequations}\label{eq:pd0}
\begin{align}
P_{i}^{d,\text{net}} &= P_{i}^{d,0}\!\left(
  a_i^{Z}\left(\frac{V_i^b}{V_i^{b,0}}\right)^{2}
 +a_i^{I}\left(\frac{V_i^b}{V_i^{b,0}}\right)
 +a_i^{P}\right),\\
Q_{i}^{d,\text{net}} &= Q_{i}^{d,0}\!\left(
  b_i^{Z}\left(\frac{V_i^b}{V_i^{b,0}}\right)^{2}
 +b_i^{I}\left(\frac{V_i^b}{V_i^{b,0}}\right)
 +b_i^{P}\right).
\end{align}    
\end{subequations}

\subsubsection{Bus Frequency Measurement}
We estimate the bus-frequency deviation at each bus \(i\in\mathcal{N}_b\), denoted \(\Delta\omega_{i}^{b}\), from the bus angle \(\delta_i^b\) using a two-stage filter: (i) a first-order low-pass with time constant $T_f$ to smoothen \(\delta_i^{b}\), followed by (ii) a high-pass filter with time constant $T_w$ that approximates the time derivative \cite{milano2010power}.
\begin{subequations}\label{eq:busfreq}
\begin{align}
  \dot{\phi}_i &= \frac{1}{T_f}\!\left(\delta_{i}^b - \phi_i\right), \label{eq:busfreq_lpf}\\[3pt]
  \dot{\psi}_i &= \frac{1}{T_w}\!\left(\phi_i - \psi_i\right), \label{eq:busfreq_wash}\\[3pt]
  \Delta\omega_{i}^b &= \frac{1}{\omega_0 T_w}\,\big(\phi_i - \psi_i\big). \label{eq:busfreq_out}
\end{align}
\end{subequations}
\subsubsection{Overall DAE Model}
We represent the overall transient DAE model as:
\begin{subequations}\label{eq:dae_system}
\begin{align}
  \dot{\mathbf{x}}_{s} &= \mathcal{F}_{s}\big(\mathbf{x}_{s},\,\mathbf{x}_{a}\big), \label{eq:dae_dyn}\\
  \mathbf{0} &= \mathcal{F}_{a}\big(\mathbf{x}_{s},\,\mathbf{x}_{a}\big). \label{eq:dae_alg}
\end{align}
\end{subequations}
where $\mathcal{F}_{s}:\mathbf{R}^{m+n}\!\to\mathbf{R}^{m}$ includes the dynamic equations (generator and governor dynamics) and $\mathcal{F}_{a}:\mathbf{R}^{m+n}\!\to\mathbf{R}^{n}$ includes the network algebraic constraints (AC power balance with ZIP loads) to define a smooth map. We define the state partitions as follows:
\begin{align*}
\mathbf{x}_{s} &= \big[\;\delta^{g},\;\Delta\omega^{g},\;P^{m}\big]^{\top} \in \mathbf{R}^{m}, \\[2pt]
\mathbf{x}_{a} &= \big[\;V^{b},\;\delta^{b}\big]^{\top} \in \mathbf{R}^{n},
\end{align*}
where $m = 3N_g$ (representing the total number of per-generator states) and $n = 2N_b$ (representing the total number of bus-voltage magnitudes and angles).

\begin{figure}[H]
    \centering
\includegraphics[width=\columnwidth,keepaspectratio]{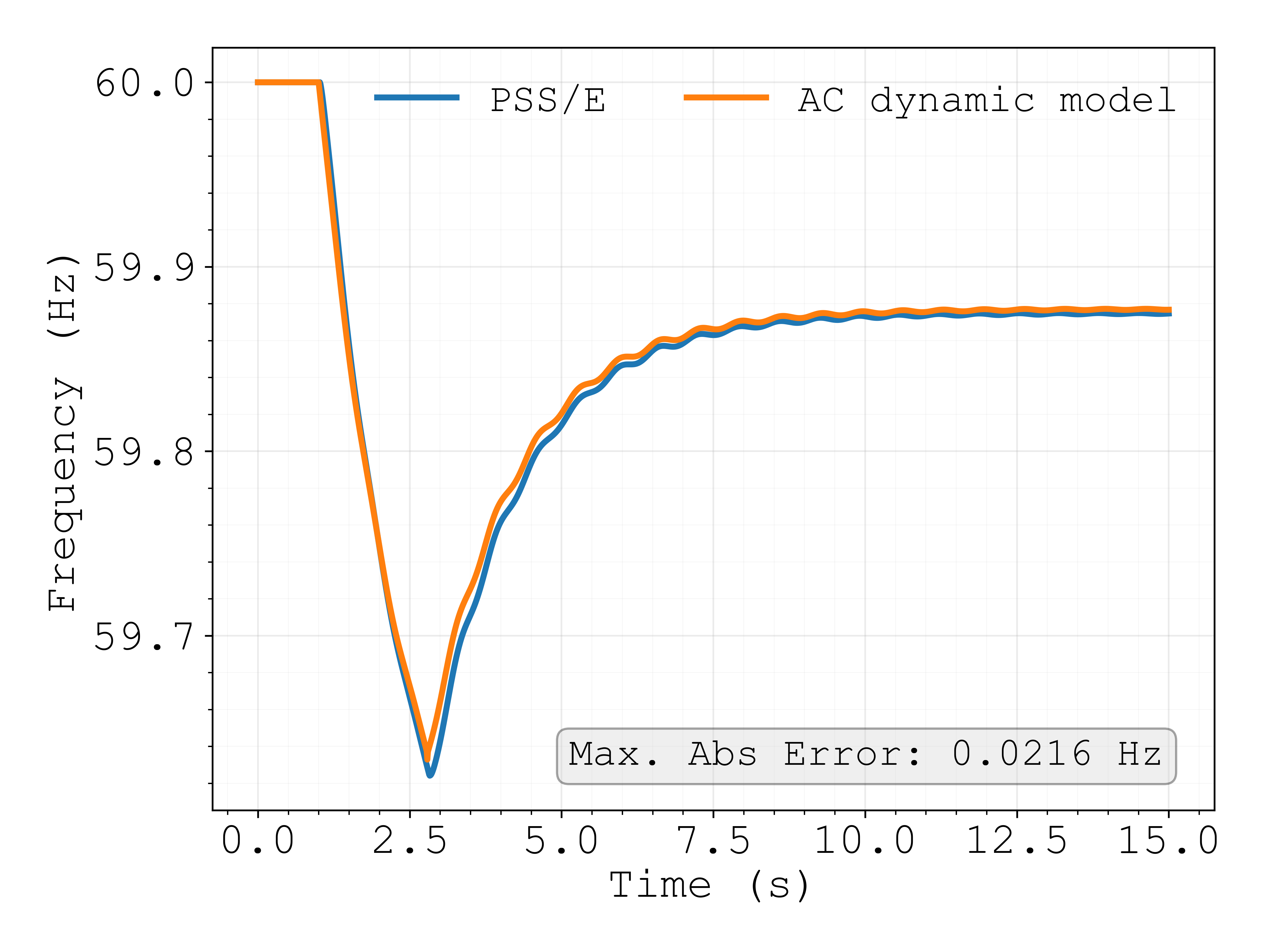}
    \caption{Validation of the AC dynamic model against PSS/E using the \textit{savnw} test case, with a 25\% generator outage introduced at 1 second.}
    \label{fig:2}
\end{figure}

We validate the accuracy of the AC DAE model in \eqref{eq:dae_system} by comparing its response against an equivalent PSS/E model. 
We compare the two models for an event that includes a load shedding action in response to a 25\% generator outage.
We implement load shedding action in PSS/E using the \verb+LDSHBL+ relay module. 
Fig.~\ref{fig:2} validates the discretized model against the PSS/E benchmark by comparing their frequency trajectories using a simulation step size of 0.01 s. The results show a maximum absolute deviation of 0.02 Hz during the transient, with both models converging to an identical steady-state frequency. This consistency proves the efficacy of the DAE model within the trajectory optimization framework.
\subsection{Droop-Based Inverter Models}
We extend the baseline transient DAE model in \eqref{eq:dae_system} to incorporate inverter-based resources (IBRs). We model the grid-following (GFL) inverter for $i \in \mathcal{N}_{\mathrm{gfl}}$ connected to a terminal bus $j \in \mathcal{N}_b$ using a simplified phase-locked loop (PLL) representation that tracks the terminal bus voltage angle $\delta_j^b$ and regulates power via frequency-watt and volt-var characteristics \cite{10136501}. For each $i \in \mathcal{N}_{\mathrm{gfl}}$, the dynamics and associated power injections are given by:

\begin{subequations}\label{eq:gfl_model}
\begin{align}
\dot{\delta}^{\mathrm{pll}}_{i} &= \omega_0 \, \Delta\omega^{\mathrm{pll}}_{i}, \label{eq:gfl_angle}\\[3pt]
\dot{\Delta\omega}^{\mathrm{pll}}_{i} &= \frac{1}{T_{i}^{\mathrm{pll}}} \left( K^{\mathrm{pll}}\big(\delta_j^{b} - \delta^{\mathrm{pll}}_i\big) - \Delta\omega^{\mathrm{pll}}_i \right), \label{eq:gfl_freq}\\[3pt]
P^{\mathrm{gfl}}_{i} &= P^{\mathrm{ref}}_{i} - k^{p}_{\mathrm{gfl}}\,\Delta\omega^{\mathrm{pll}}_{i}, \label{eq:gfl_peq} \\[3pt]
Q^{\mathrm{gfl}}_{i} &= Q^{\mathrm{ref}}_{i} - k^{q}_{\mathrm{gfl}}\left(V_j^{b} - V_j^{b,0}\right). \label{eq:gfl_pq}
\end{align}
\end{subequations}

\noindent where $T_i^{\mathrm{pll}}$ and $K^{\mathrm{pll}}$ are the PLL time constant and gain, while $k^p_{\mathrm{gfl}}$ and $k^q_{\mathrm{gfl}}$ represent the active and reactive power droop coefficients, respectively. While grid-following (GFL) inverters primarily track the existing grid frequency to regulate power injections, grid-forming (GFM) inverters actively establish the local grid frequency and voltage through droop-based synchronization. For $i \in \mathcal{N}_{\mathrm{gfm}}$, we use a third-order model that captures the active power-frequency ($P-\Delta \omega$) and reactive power-voltage ($Q-E$) droop dynamics \cite{du2021model}. To distinguish these from synchronous machines, we denote the inverter internal voltage magnitude as $E_{i}^{\mathrm{d}}$, the internal angle as $\delta_{i}^{\mathrm{d}}$, and the internal frequency deviation as $\Delta\omega_{i}^{\mathrm{d}}$, governed by:

\begin{subequations}\label{eq:gfm_model}
\begin{align}
\dot{\delta}_i^{\mathrm{d}} &= \omega_0\,\Delta \omega^{\mathrm{d}}_i, \label{eq:gfm_delta_dot}\\[3pt]
\dot{\Delta \omega}_i^{\mathrm{d}} &= \frac{1}{T_{i}^{\mathrm{w}}}\Big( k^{p}_{\mathrm{gfm}}(P_i^{\mathrm{ref}} - P_i^{\mathrm{inv}}) - \Delta \omega_i^d \Big), \label{eq:gfm_omega_dot}\\[3pt]
\dot{E}^{\mathrm{d}}_i &= \frac{1}{T_{i}^{\mathrm{w}}}\Big(k^{q}_{\mathrm{gfm}} (Q_{i}^{\mathrm{ref}} - Q_{i}^{\mathrm{inv}}) - (E^{\mathrm{d}}_i - E_i^0 )  \Big). \label{eq:gfm_e_dot}
\end{align}
\end{subequations}

\noindent The active and reactive power injections, $P_{i}^{\mathrm{inv}}$ and $Q_{i}^{\mathrm{inv}}$, are determined by the algebraic coupling with the terminal bus $j \in \mathcal{N}_b$ as:
\begin{subequations}\label{eq:gfm_peq}
\begin{align}
P_{i}^{\mathrm{inv}} &= \frac{E^{\mathrm{d}}_i\,V_{j}^{b}}{X_{i}^{\mathrm{d}}} \,\sin(\delta_{i}^{\mathrm{d}} - \delta_{j}^{b}), \label{eq:gfm_Pe}\\[3pt]
Q_{i}^{\mathrm{inv}} &= \frac{E^{\mathrm{d}}_i\,V_{j}^{b}}{X_{i}^{\mathrm{d}}} \,\cos(\delta_{i}^{\mathrm{d}} - \delta_{j}^{b}) \;-\; \frac{(V_{j}^{b})^2}{X_{i}^{\mathrm{d}}}. \label{eq:gfm_Qe}
\end{align}
\end{subequations}

\noindent where $X_{i}^{\mathrm{d}}$ represents the inverter coupling reactance. Here, $T_i^{\mathrm{w}}$ is the power measurement filter time constant, while $k^p_{\mathrm{gfm}}$ and $k^q_{\mathrm{gfm}}$ are the frequency and voltage droop gains, respectively. We initialize the reference internal voltage $E_i^0$ to match the pre-disturbance steady-state operating point as detailed in the initialization setup.
\section{Adaptive UFLS Design: A Trajectory  Optimization Approach}\label{sec:prob_for}
In the adaptive UFLS design problem, we compute the optimal relay frequency setpoints at each scheduling period $p$, whose interval depends on the system operator. 
We design the framework assuming the period $p$ is an hour, as the net variations in load and inertia within this interval remain limited under normal operating conditions. The optimization ensures stability under a NERC-defined worst-case disturbance\footnote{Following NERC PRC-006-2 criteria, we define the worst-case disturbance as a significant generation-load imbalance, typically resulting from the instantaneous loss of approximately 25\% of the total system generation capacity~\cite{NERC_PRC006_2}.} and can be rerun earlier if system conditions change significantly. 
To determine the optimal frequency setpoints, we solve a \tos{} problem that determines the percentage of load to shed at each stage and the location of action for the UFLS relay. 
\TO{} computes an optimal solution by identifying a sequence of admissible controls\footnote{The set of control inputs that generate a feasible trajectory satisfying all system constraints. In the adaptive UFLS problem, the control actions are relay switching that guides a dynamic system along a feasible path while optimizing a given objective function \cite{kelly2017introduction}}. \TO{} explicitly embeds generator swing equations, governor–turbine dynamics, and AC power-flow constraints into the optimization, so that load-shedding actions enforce the governing differential–algebraic equations and yield optimal trajectories.

Many approaches exist to formulate a \tos{} problem. For the adaptive UFLS design problem, we use a continuous trajectory formulation, where the system dynamics remain continuous throughout the entire trajectory. There are two main types of continuous \tos{} methods: \textit{direct} and \textit{indirect}. The direct method discretizes the continuous-time differential constraint set into an algebraic constraint set, a process known as direct transcription. In contrast, the indirect method derives analytical expressions for satisfying necessary conditions for optimality. 
In this work, we employ the trapezoidal collocation method, a direct transcription method to minimize the load shed to enforce stable frequency response during a NERC-specified worst-case contingency\cite{NERC_PRC006_2}.
We implement the design to be a staged mechanism in which portions of the load disconnect once measured frequency values fall below certain thresholds. 

We recognize and consider that certain system loads are inherently heterogeneous and have clear priorities. 
Critical loads, such as hospitals and elder care facilities, are therefore last to disconnect in this staged UFLS design. 
We achieve this by partitioning the load into three representative stages: the first stage includes residential demand, the second stage encompasses commercial demand, and the third stage addresses the most critical loads. We define $z_{i}^\text{agg}[k]$ in \eqref{eq:switch} as the sum of total fraction of load shed at bus $i$ during time step $k$. This aggregate value sums the binary decision variables $z_i^s[k]$, which determine the switching status of each load stage $s \in \mathcal{S}$, weighted by their respective prescribed shedding portions $\alpha_i^s$.  
\begin{equation}\label{eq:switch}
    z_{i}^\text{agg}[k] = \sum_{s \in \mathcal{S}} \alpha_i^s z_i^s[k],
\end{equation}  
Subsequently, \eqref{eq:ac_pf} is updated by adding the UFLS actions:
\begin{align}
    P_{i}^{d}[k] &= (1-z_{i}^\text{agg}[k]) (P_{i}[k])^{d,\text{net}}\label{eq:UFLS_actions_1} \\
    Q_{i}^{d}[k] &= (1-z_{i}^\text{agg}[k]) (Q_{i}[k])^{d,\text{net}} \label{eq:UFLS_actions_2}
\end{align}
The overall \tos{} is defined in \eqref{eq:problem_definition}.
The objective function in~\eqref{eq:obj} minimizes the total amount of load shed, where $P^{\mathrm{d,0}}_{i}$ denotes the nominal active power demand at load bus $i$ from \eqref{eq:pd0}. Constraint~\eqref{eq:mono} enforces the \emph{monotonicity of stage activations over time}. Once a shedding stage at bus $i$ is triggered at step $k\!-\!1$, the corresponding binary variable must remain active for all subsequent steps $k$ through $N$. This prevents reclosing of loads within the optimization trajectory. Constraints ~\eqref{eq:cascade_1} and~\eqref{eq:cascade_2} enforce the \emph{stage prioritization and delay}. For each bus $i$, a higher-order stage $s$ can only be activated if the preceding stage $s\!-\!1$ has already been triggered. To introduce an inter-stage delay, we define the event detector in~\eqref{eq:cascade_1} as the discrete-time derivative of the shedding variable. Because of the monotonicity condition in~\eqref{eq:mono}, the detector has value 1 only at time-step $k$ when the switch flips from 0 to 1. 
We then enforce a strict temporal separation of $\Delta d$ steps between the activation events of consecutive stages in~\eqref{eq:cascade_2} as specified in NERC criteria \cite{NERC_PRC006_2}. The choice of $\Delta d$ depends on the time delay and choice of step size $h$. 
We compute $\Delta d =  \cfrac{\tau^s}{h}$, where $\tau^s$ is the desired inter-stage delay time for each stage.

\begin{subequations}\label{eq:problem_definition}
\setlength\parindent{0pt}\rule{\columnwidth}{1pt}
\textbf{Adaptive UFLS Design Trajectory Optimization} \\\rule[1ex]{\columnwidth}{1pt}\vspace*{-1.5\baselineskip}
\begin{small}
\begin{alignat}{2}
    & \qquad \quad \min_{z, x} \sum_{k=0}^{N}\sum_{i\in\mathcal{N}_l}\sum_{s\in\mathcal{S}}
      \alpha_{i}^{s}\, z_{i}^{s}[k]\; P^{d,0}_{i} \label{eq:obj} \\[-0.4ex]
     \shortintertext{\text{s.t.}}
     \shortintertext{$\forall i \in \mathcal{N}_l,\; s \in \mathcal{S}$}
    \shortintertext{\underline{Monotonicity Constraint}}
    & z_{i}^{s}[k] \ge z_{i}^{s}[k-1], \; \forall \; k=1,\dots,N\label{eq:mono} \\
    \shortintertext{\underline{Stage Prioritization and Delay Constraints}}
    & y_{i}^{s}[k] = z_{i}^{s}[k] - z_{i}^s[k-1],\; \forall \; k=1,\dots,N\label{eq:cascade_1} \\
    & y_{i}^{s}[k] \leq y_{i}^{s-1}[k-\Delta d],\; \forall \; s\in\mathcal{S}\setminus\{1\}, k=\Delta d,\dots,N \label{eq:cascade_2}\\
    \shortintertext{\underline{Settling Frequency}}
    & \underline{\Delta \omega^{ss}} \le \Delta\omega_{i}^{b}[N] \le \overline{\Delta \omega^{ss}}\label{eq:freqband} \\
    \shortintertext{\underline{System Dynamics}}
    & \mathbf{x_s}[k+1] = \mathbf{x_s}[k] + \frac{h}{2} \Big[ \mathcal{F}_{s}\big(\mathbf{x_s}[k+1], \mathbf{x_a}[k+1], \nonumber \\
    & \hspace{5.5em} \mathbf{z}[k+1]\big) + \mathcal{F}_{s}\big(\mathbf{x_s}[k], \mathbf{x_a}[k], \mathbf{z}[k]\big) \Big], \nonumber \\
    & \hspace{5.5em} \forall \; k = 0,\dots,N-1 \label{eq:dyn_discrete} \\
    & \mathbf{0} = \mathcal{F}_{a}\big(\mathbf{x_s}[k], \mathbf{x_a}[k], \mathbf{z}[k]\big), \forall \; k = 0,\dots,N \label{eq:alg_discrete} \\
    \shortintertext{\underline{Variable Bounds and Boundary conditions}}
    & z_{i}^{s}[k] \in \{0,1\}, \forall \; k = 0,\dots,N\\
    & \Delta\omega_{i}^{b}[k] \geq \underline{\Delta \omega}, \forall \; k = 0,\dots,N
    \label{eq:z_box} \\
    & \mathbf{x_s}[0] = \mathbf{x_s}^{0} 
\end{alignat}
\end{small}
\rule[1ex]{\columnwidth}{1pt}
\end{subequations}

The formulation \eqref{eq:problem_definition} considers the practical design needs of UFLS schemes, where shedding is executed sequentially; less critical loads are given priority to be shed.
Further, adding a small time delay $\tau^s$ prevents multiple stages from activating simultaneously due to small oscillations or measurement noise.  The constraint~\eqref{eq:freqband} ensures that settling bus frequencies remain within a safe operational band, thereby maintaining system stability and avoiding excessive load disconnections. The constraint~\eqref{eq:z_box} ensures all frequencies are above the acceptable frequency nadir for all the time steps.  
The constraints~\eqref{eq:dyn_discrete} and~\eqref{eq:alg_discrete} model the AC grid dynamics (discussed in Section-\ref{sec:prelim}). 
The effective active and reactive load demands defined in \eqref{eq:UFLS_actions_1}, \eqref{eq:UFLS_actions_2} with $z_{i}^\text{agg}[k]$ given by \eqref{eq:switch} enter the algebraic equations $\mathcal{F}_a(.)$ in \eqref{eq:alg_discrete}, thereby coupling the binary stage variables to the network power-flow model.
In~\eqref{eq:dyn_discrete} and~\eqref{eq:alg_discrete}, $\mathbf{z}[k]$ is the vector that collects the on/off statuses of all load-shedding relays across load buses ($\mathcal{N}_l$) and stages ($\mathcal{S}$) at time step k. Constraints \eqref{eq:dyn_discrete} and \eqref{eq:alg_discrete} enforce that the variables follow the transient trajectory of a specific NERC-defined worst-case disturbance, thereby tuning the load-shedding decision variables for that single event. The formulation can be extended to multiple disturbance scenarios, although this would significantly increase the number of optimization variables and the overall problem size.

\noindent \textit{Initialization:} We preprocess the network to define the boundary conditions and allocate the load-shedding decision variables. 
We impose boundary conditions using the steady-state operating point. Specifically, 
$\mathbf{x}_{s}[0] = \mathbf{x}_{s}^{0}$, 
$\mathbf{x}_{a}[0] = \mathbf{x}_{a}^{0}$, and 
$z_{i}^{s}[0] = 0$ for all $i \in \mathcal{N}_l$ and $s \in \mathcal{S}$. 
We obtain the algebraic variables $\mathbf{x}_{a}^{0} = [\,V_i^b,\,\delta_i^b\,]^\top$ by solving the power-flow equations defined in \eqref{eq:ac_pf}. 
We initialize the dynamic states of every generator $i \in \mathcal{N}_g$, by setting the mechanical power, $P_i^{m}[0]$, to PV bus power flow input $P_i^{g}$ and we set the initial frequency deviation, $\Delta \omega_i^{g}[0]$, to  $0$. 
We then compute the generator internal states $(E_i^{\prime},\,\delta_i^{g})$ from the power-flow solution. 
Given the bus phasor voltage $\tilde{V}_i^b = V_i^b e^{j\delta_i^b}$, the reactance $Z_i = jX_i^{d\prime}$, and the apparent power injection $S_i^g = P_i^g + jQ_i^g$, we calculate the following: 
\begin{subequations}\label{eq:init_gen_states}
\begin{align}
    I_i^g &= \frac{(S_i^g)^{*}}{(\tilde{V}_i^b)^{*}}, 
    &\quad E_i &= \tilde{V}_i^b + I_i^g Z_i, \label{eq:init_voltage}\\
    E_i^\prime &= |E_i|, 
    &\quad \delta_i^g &= \angle E_i. \label{eq:init_delta}
\end{align}
\end{subequations}
Since certain buses with distributed energy resources (DERs) may backfeed power into the grid during the study period $p$, we address these cases in the preprocessing stage (see Remark \ref{re:1}).
\begin{remark}
\label{re:1}
We identify backfeeding buses during preprocessing and exclude them from the set of candidate load-shedding locations. This ensures that the formulation does not assign shedding actions to nodes injecting power into the grid
\end{remark}

The trajectory optimization formulation in \eqref{eq:problem_definition} identifies the minimum amount of load shedding across locations.
The framework post-processes the results to obtain practical UFLS relay settings. In this postprocessing stage, we extract the system frequencies from the optimal solution at which the relay at various locations and stages triggers shedding. 
We then assign these extracted frequencies as the trigger thresholds for the corresponding relay stages.
We update these settings at every predetermined operator-defined time period.
By updating relay settings recursively in time, the approach overcomes the drawbacks of conventional UFLS. Nonetheless, the problem is \textbf{NP}-hard MINLP.

\section{{Homotopy Driven Integer Enforcement Framework}}\label{sec:Integer_enforce}
\begin{figure*}[]
    \centering
    \begin{subfigure}{0.32\textwidth}
        \centering
        \includegraphics[width=\linewidth]{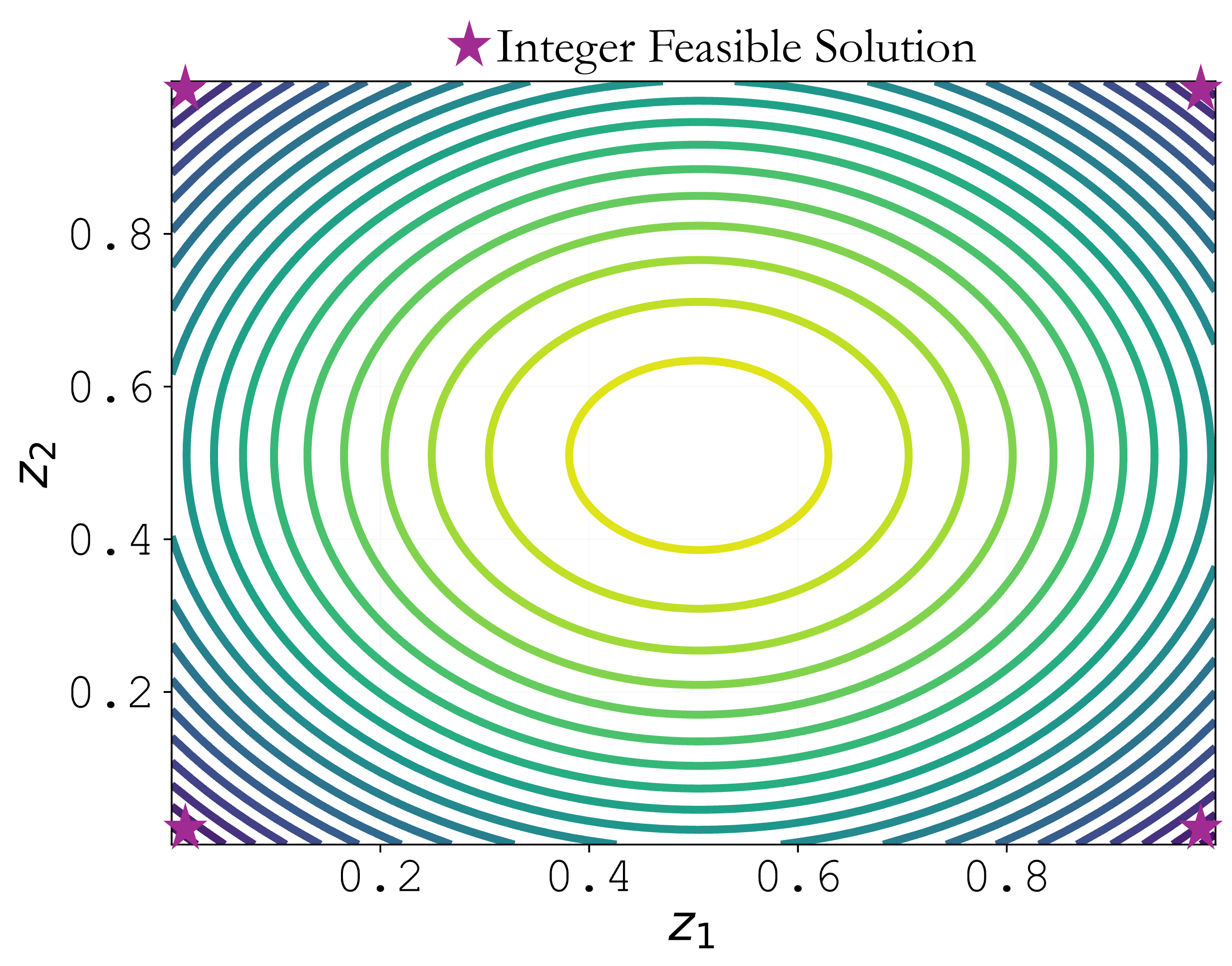}
        \caption[]{} 
        \label{fig:original_prob}
    \end{subfigure}
    \hfill 
    \begin{subfigure}{0.32\textwidth}
        \centering
        \includegraphics[width=\linewidth]{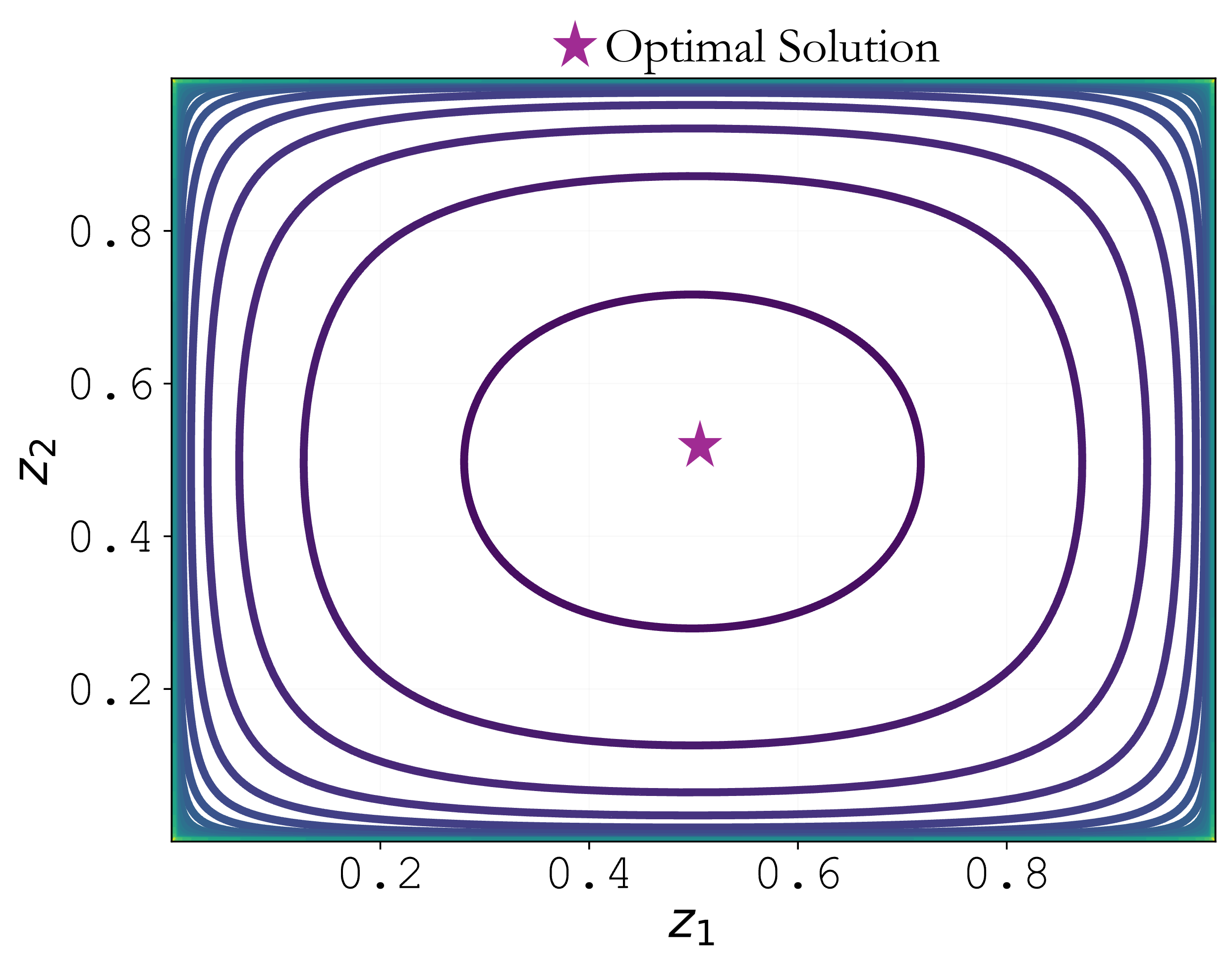}
        \caption[]{} 
        \label{fig:perturbed_prob}
    \end{subfigure}
    \hfill 
    \begin{subfigure}{0.32\textwidth}
        \centering
        \includegraphics[width=\linewidth]{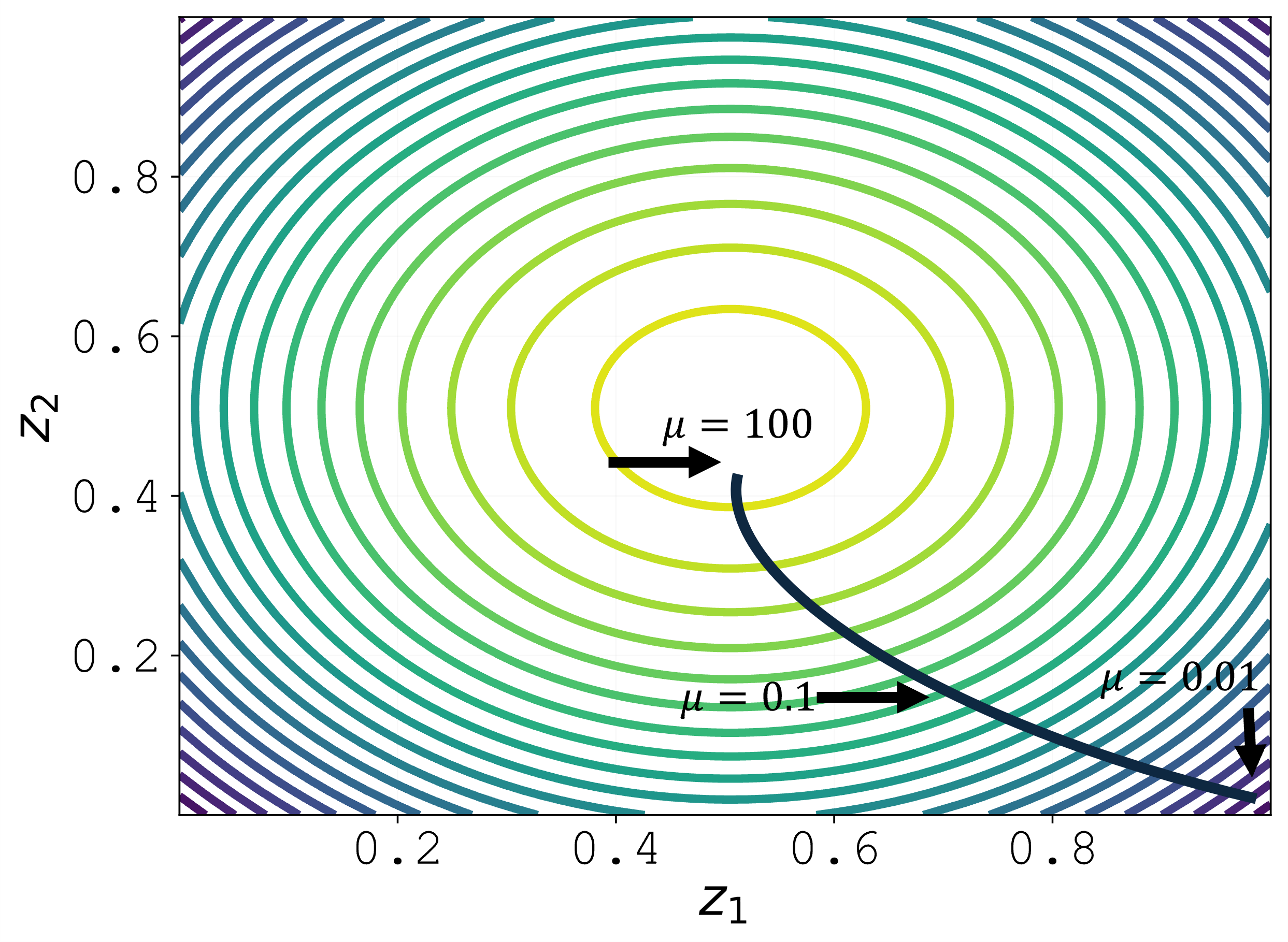}
        \caption[]{} 
        \label{fig:Trajectory_sol}
    \end{subfigure}
\caption{(a) Contour plot for original problem $\mathbf{P}_3$ objective with quadratic penalty (b) Perturbed \textit{convex} objective after introducing the barrier term with the value of $\mu > \gamma$  (c) Solution trajectory with the decreasing values of the homotopy parameter $\mu$. As $\mu \to 0$, the function's minimum is driven to a vertex of the original feasible region, yielding an integer feasible solution.}
    \label{fig:homotopy_all}
\end{figure*}
The adaptive UFLS design \tos{} in~\eqref{eq:problem_definition} is a MINLP because the constraint set includes the binary relay variables and nonlinear AC network dynamics. 
We build a homotopy-driven framework to obtain an AC-feasible solution that satisfies stationarity conditions and gives provably near-integer feasible solutions for binary variables.
We represent the original MINLP \eqref{eq:problem_definition} in a compact form with~\eqref{P1:minlp}. In ~\eqref{P1:minlp}, the objective function~\eqref{eq:P1_gen_obj} mimics the objective in~\eqref{eq:obj}. The equality constraints~\eqref{eq:P1_gen_eq} consist of collocation constraints for  system dynamics~\eqref{eq:dyn_discrete} and~\eqref{eq:alg_discrete}, while the inequality constraints~\eqref{eq:P1_gen_ineq} consist of~\eqref{eq:mono},~\eqref{eq:cascade_1},~\eqref{eq:cascade_2},~\eqref{eq:freqband} and~\eqref{eq:z_box}. We use $n_z$ to denote the number of binary decision variables associated with load-shedding actions in \eqref{eq:P1_gen_domains}, and $n_x$ to denote the number of continuous algebraic and dynamic state variables for each time step $k$.

\begin{subequations}\label{P1:minlp}
\setlength\parindent{0pt}\rule{\columnwidth}{1pt}
\textbf{$\mathbf{P}_{1}$: MINLP Formulation} \\\rule[1ex]{\columnwidth}{1pt}\vspace*{-1.5\baselineskip}
\begin{small}
\begin{alignat}{2}
    \quad & \min_{x,z} \sum_{k=0}^{N} F(x[k], z[k]) \label{eq:P1_gen_obj} \\
    & \text{s.t.} \\ 
    & \quad G(x[k], z[k]) = 0,\; k=0,\dots,N, \label{eq:P1_gen_eq}\\
    & \quad H(x[k], z[k]) \le 0, \; k=0,\dots,N, \label{eq:P1_gen_ineq}\\
    & \quad x[k] \in \mathbb{R}^{n_x}, \quad \colorbox{gray!18}{$ z[k] \in \{0,1\}^{n_z}$}, \; k=0,\dots,N. \label{eq:P1_gen_domains}
\end{alignat}
\end{small}
\vspace{-0.3em}
{\footnotesize%
\textit{Note:} $z[k]$ denotes the stage-wise binary decision as highlighted.
}
\rule[1ex]{\columnwidth}{1pt}
\end{subequations}

Various solvers, such as Knitro, Bonmin, BARON, and Couenne, are available to solve MINLP problems. While these solvers utilize various algorithmic approaches, they are generally reliable only for convex mixed-integer programs (see \cite{kronqvist2019review}) and have shown limited effectiveness when applied to large-scale problems \cite{leyffer2009applications}-\cite{murray2010algorithm}. 

Several other approaches solve MINLP problems by relaxing binary variables to continuous ones. 
Most trivial amongst them treats binary variables as continuous and discretizes the solution through rounding; however, this can compromise feasibility under certain conditions \cite{gill2019practical}. Alternatively, regularization schemes can more robustly relax MINLPs into NLPs by introducing a parametric constraint of the form $z^T (\hat{e}-z)\leq \tau $, where $0 \leq z\leq 1$, and $\hat{e}$ is the vector of all ones. 
This constraint resembles the complementarity KKT condition, and by choosing a small positive value ($\tau \geq 0$), the constraint pushes the value $z$ to be either near 0 or 1. This regularization technique was first introduced in \cite{scholtes2001convergence} and is commonly applied to a class of problems known as mathematical programs with equilibrium constraints (MPECs). 

In our case, however, we found that using very small values of $\tau$ often rendered the problem infeasible, which limited the practicality of applying this method directly to the adaptive UFLS problem.
An alternative regularization approach involves replacing the complementarity condition-like constraint with a family of functions called \textit{NCP-}functions \cite{fischer1992special}, then applying regularization to it. These all regularization schemes \cite{scholtes2001convergence}-\cite{fischer1992special} work by effectively tightening the regularization parameter $\tau$, and each tightening step generates a disjoint feasible set leading to convergence issues, as the NLP solver may jump between disconnected regions of the feasible space. 

Rather than including the complementarity constraint to enforce binary behavior, in our approach in $\textbf{P}_2$, we add the penalty term to the objective with penalty parameter ($\gamma$) \cite{borchardt1988exact}. 
We posit that the penalty method is closer to the strongest form of regularization, since its feasible set is a superset of that of the complementarity-based constraint set \cite{schewe2019computing}. 
Moreover, results in \cite{murray2010algorithm} show that, under mild assumptions, the penalized formulation serves as an exact reformulation of the original MINLP when $\gamma$ is chosen sufficiently large.

The equivalent formulation of the problem \eqref{P1:minlp} is described in 
\eqref{P2:nlp}, denoted as $\mathbf{P}_2$. The quadratic penalty in \eqref{eq:P2_gen_obj_2} is zero exactly at $z \in \{0,1\}$ and positive otherwise; thus, increasing $\gamma$ promotes an integer-feasible solution in $\mathbf{P}_2$. One potential strategy involves the iterative tightening of $\gamma$ to drive variables toward the integer feasible solution. 

\begin{subequations}\label{P2:nlp}
\setlength\parindent{0pt}\rule{\columnwidth}{1pt}
\textbf{$\mathbf{P}_{2}$: NLP with Quadratic Penalty} \\\rule[1ex]
{\columnwidth}{1pt}\vspace*{-1.5\baselineskip}
\begin{small}
\begin{align}
   \quad & \min_{x,z} \sum_{k=0}^{N} F(x[k], z[k]) + \colorbox{gray!18}{$\gamma \sum_{k=0}^{N} z[k] (1-z[k])$}\label{eq:P2_gen_obj_2} \\
    & \text{s.t.} \\  
    & \quad G(x[k], z[k]) = 0,\; k=0,\dots,N, \label{eq:P2_gen_eq_2}\\
    & \quad H(x[k], z[k]) \le 0, \; k=0,\dots,N, \label{eq:P2_gen_ineq_2}\\
    & \quad x[k] \in \mathbb{R}^{n_x}, \quad \colorbox{gray!18}{$z[k] \in [0,1]^{n_z}$}, \; k=0,\dots,N. \label{eq:P2_gen_domains}
\end{align}
\end{small}
\vspace{-0.3em}
{\scriptsize%
\textit{Note:} $z[k]$ is continuous, with an additional penalty term in the objective as highlighted.
}
\rule[1ex]{\columnwidth}{1pt}
\end{subequations}

However, this penalty makes the objective function concave in $z$, which introduces an exponential number of local minima at the feasible region's vertices. We empirically observed that relying solely on the iterative tightening of $\gamma$ often forces the solver to converge prematurely to a nearest local minima determined by the initial starting point. This approach frequently leads to convergence failure or poor-quality solutions in large-scale systems because the solver lacks a guiding trajectory, as also highlighted in \cite{murray2010algorithm}.

\begin{subequations}\label{P3:nlp}
\setlength\parindent{0pt}\rule{\columnwidth}{1pt}
\textbf{$\mathbf{P}_{3}$: Regularized NLP} \\\rule[1ex]
{\columnwidth}{1pt}\vspace*{-1.5\baselineskip}
\begin{small}
\begin{align}
    & \min_{x,z} 
      \begin{aligned}[t]
        & \sum_{k=0}^{N} F(x[k], z[k])+ \gamma \sum_{k=0}^{N} z[k] (1-z[k]) \\
        &\colorbox{gray!18}{$-\mu \sum_{k=0}^{N} (ln(z[k]) + ln(1-z[k]))$}
      \end{aligned} \label{eq:P3_gen_obj_3} \\[1ex]
    & \text{s.t.} \\
    & \quad G(x[k], z[k]) = 0,\; k=0,\dots,N, \label{eq:P3_gen_eq_3}\\
    & \quad H(x[k], z[k]) \le 0,\; k=0,\dots,N, \label{eq:P3_gen_ineq_2}\\
    & \quad x[k] \in \mathbb{R}^{n_x}, \quad \colorbox{gray!18}{$z[k] \in [0,1]^{n_z}$}, \; k=0,\dots,N. \label{eq:P3_gen_domains}
\end{align}
\end{small}
\vspace{-0.3em}
{\scriptsize%
\textit{Note:} $z[k]$ is continuous, with an added homotopy term in the objective as highlighted.
}
\rule[1ex]{\columnwidth}{1pt}
\end{subequations}

To overcome this, we apply a homotopy-based strategy that combines a quadratic penalty with a logarithmic barrier. In $\mathbf{P}_3$ we adopt the smoothing approach of \cite{murray2010algorithm} and introduce the barrier term $-\mu(\ln z + \ln(1-z))$, where $\mu$ serves as a homotopy parameter as highlighted in \eqref{eq:P3_gen_obj_3}. The barrier term is strictly convex with $z>0$. By making the initial objective function convex (see Remark \ref{re:2}), we create a surrogate problem with a clear, well-defined local minimum that simplifies the search in this complex region. This strategy enforces iterates of $z$ to remain strictly inside $(0,1)$, starting with large $\mu$. Here, the small $\gamma$ produces a smooth interior relaxation.
Now, during the homotopy trajectory, gradually decreasing $\mu$ while increasing $\gamma$ drives the iterates toward $\{0,1\}$ without causing discontinuous changes in the feasible set. See algorithm \ref{alg:homotopy} for details.

\begin{remark}
\label{re:2}
The convex surrogate objective 
$-\mu \,[\ln(z)+\ln(1-z)]$ regularizes the relaxation of binary decisions, provides a desirable starting point around 0.5 for subsequent iterations, and helps the NLP solver to converge more reliably.
\end{remark}

We formulate the problem in~\eqref{P3:nlp} by adding a barrier term, where $\mu$ acts as the homotopy parameter and $\gamma$ as the penalty parameter. Fig. \ref{fig:homotopy_all} illustrates the main idea of the algorithm. In Fig. \ref{fig:homotopy_all}(a), the quadratic penalty term attains its maximum at $z=0.5$, while in Fig. \ref{fig:homotopy_all}(b), the perturbed problem with the barrier term reaches its minimum at the same point. This structure creates a natural interior starting position around $z=0.5$. As we gradually decrease $\mu$ and increase $\gamma$, the formulation defines a continuous trajectory that guides the iterates toward the boundary, ultimately reaching an integer-feasible solution. The overall algorithm is presented in Algorithm~\ref {alg:homotopy}. We update the parameters, $\mu$ and $\gamma$ using the strategy proposed in \cite{murray2010algorithm}, we set the decay factor $\rho_\mu$ 
for $\mu$ in the range $[0.1,\,0.9]$ and the growth factor $\rho_\gamma > 1$ with the product $\rho_\mu\rho_\gamma \approx 1$. In practice, we tune $\rho_\mu$ and 
$\rho_\gamma$ in a manner similar to hyperparameter tuning.  
\begin{algorithm}[t]
\caption{UFLS Trajectory Optimization (\textbf{P3})}
\label{alg:homotopy}
\KwInput{Steady-state data $\mathcal{D}_{\mathrm{ss}}$, dynamics data $\mathcal{D}_{\mathrm{dyn}}$, total steps $N$, step size $h$, tolerance $\varepsilon>0$}
\KwOutput{Optimized trajectory $\{x[k]^\star\}_{k=0}^{N}$, where $x=\{x_s,x_a\}$, switch status $\{z_i^s[k]^\star\}_{k=0}^{N}$, and frequency thresholds $\Delta \omega^{\mathrm{th}}_{i,s}$ where, $i\in \mathcal{N}_l$, $s\in \mathcal{S}$}

\While{adaptive UFLS in operation:}{
  \Comment*{\scriptsize repeat every period $p$}
  \textbf{Parse data:} $\mathcal{D}_{\mathrm{ss}}, \mathcal{D}_{\mathrm{dyn}}$\;
  \textbf{Initialization:} solve for steady state to obtain $x[0] \gets x_{\text{init}}$\;
  \textbf{Parameter selection:} choose $\mu_0 \ge 0$, $\gamma_0 \ge 0$ with $\mu_0 > \gamma_0$\;
  \textbf{Epoch and rates:} set $e \gets 0$; choose $\rho_\mu \in (0.1,0.9)$, $\rho_\gamma > 1$ with $\rho_\mu \rho_\gamma = 1$\;
  \While{$(e=0)\;\lor\;\bigl(\exists\,k:\ z_i^s[k]^{(e)}(1-z_i^s[k]^{(e)}) > \varepsilon \bigr)$}{
    \If{$e > 0$}{
      \textbf{Warm start:} $x[k] \gets x[k]^{(e)}$, $z_i^s[k] \gets z_i^s[k]^{(e)}$\;
      \Comment*{\scriptsize for all $i\in\mathcal{N}_l,\; s\in\mathcal{S},\; k=0,\dots,N$}
    }
    \textbf{Solve} $\mathbf{P}_3$ with $(\mu_e,\gamma_e)$ to get $x[k]^{(e+1)}, z_i^s[k]^{(e+1)}$\;
    \Comment*{\scriptsize for all $i\in\mathcal{N}_l,\; s\in\mathcal{S},\; k=0,\dots,N$}
    \textbf{Update parameters:} $\mu_{e+1} \gets \rho_\mu \mu_e$, \quad $\gamma_{e+1} \gets \rho_\gamma \gamma_e$\;
    $e \gets e+1$
  }
  \textbf{Collect:} $\{x[k]^\star, z_i^s[k]^\star\} \gets \{x[k]^{(e)}, z_i^s[k]^{(e)}\}$; compute final trajectory for $k=0,\dots,N$\;
  \textbf{Post processing:} For each $(i,s)$, let 
  $k^\mathrm{act}_{i,s}= \operatorname*{arg\,min}_{k}\{k \mid z_i^s[k]^\star \ge 0.9\}$; 
  set $\Delta \omega^{\mathrm{th}}_{i,s}\gets \Delta\omega_{i}^{b}[k^\mathrm{act}_{i,s}]^\star$
}
\end{algorithm}

\section{Results and Discussion}\label{sec:res}
We validate the proposed framework by implementing Algorithm~\ref{alg:homotopy} and conducting simulations on synthetic power-system test cases obtained from \cite{tamu_electric_grid_repository}. Detailed simulation parameters are provided in the Appendix. We perform all experiments on a laptop computer equipped with an AMD Ryzen 9 processor and 32~GB RAM, running Ubuntu Linux~24.04 under the Windows Subsystem for Linux (WSL~2). We solve the resulting NLPs using Ipopt~(version~3.14.16) with the HSL~MA57 linear solver \cite{hsl2025}.
\subsection{Description of Test Cases}
We first employ the WSCC 9-bus system, a simple approximation of the Western System Coordinating Council with three generators and three loads discussed in Fig.~\ref{fig:methodology}. We then analyze the \texttt{savnw} test case from PSS/E, a 23-bus, three-area, six-machine system.
Our study also includes the IEEE 59-bus network with 33 generators and 19 loads, along with the ACTIVSg200 and ACTIVSg500 synthetic grids, which represent the footprints of central Illinois and South Carolina, respectively, and the \texttt{bench} example network extracted from PSS/E, consisting of 1648 nodes with 313 generators and 1220 loads.
In all test cases, we consider the peak-load operating condition. To create the disturbance, we simulate a 25\% generation loss at $t=1$ s by removing a generator or a set of generators whose combined capacity accounts for approximately 25\% of the total generation, which aligns with the NERC PRC-006-2 worst-case disturbance criteria \cite{NERC_PRC006_2}. We use a sampling time of 0.1 s for \texttt{savnw} systems, which is used for the discussion of results in Section~\ref{sub:result_1} and Section~\ref{subsec:results_2}. For the larger networks (ACTIVSg200, ACTIVSg500, and \texttt{bench}), we increase the sampling time to 0.5 s to reduce the computational burden.  

We ensure consistency with the mathematical dynamical model presented in Section-\ref{sec:prelim} by replacing the existing dynamic data with classical generator models and TGOV governor models. For buses without dynamic model specifications, we added negative-load injections to represent generation. We provide the complete modified test cases at \url{https://github.com/hamzaali412/Dynamic-Test-Cases.git}. 

\subsection{Practical Adaptive UFLS Design}\label{sub:result_1}
\begin{figure}
    \centering
\includegraphics[width=\columnwidth,keepaspectratio]{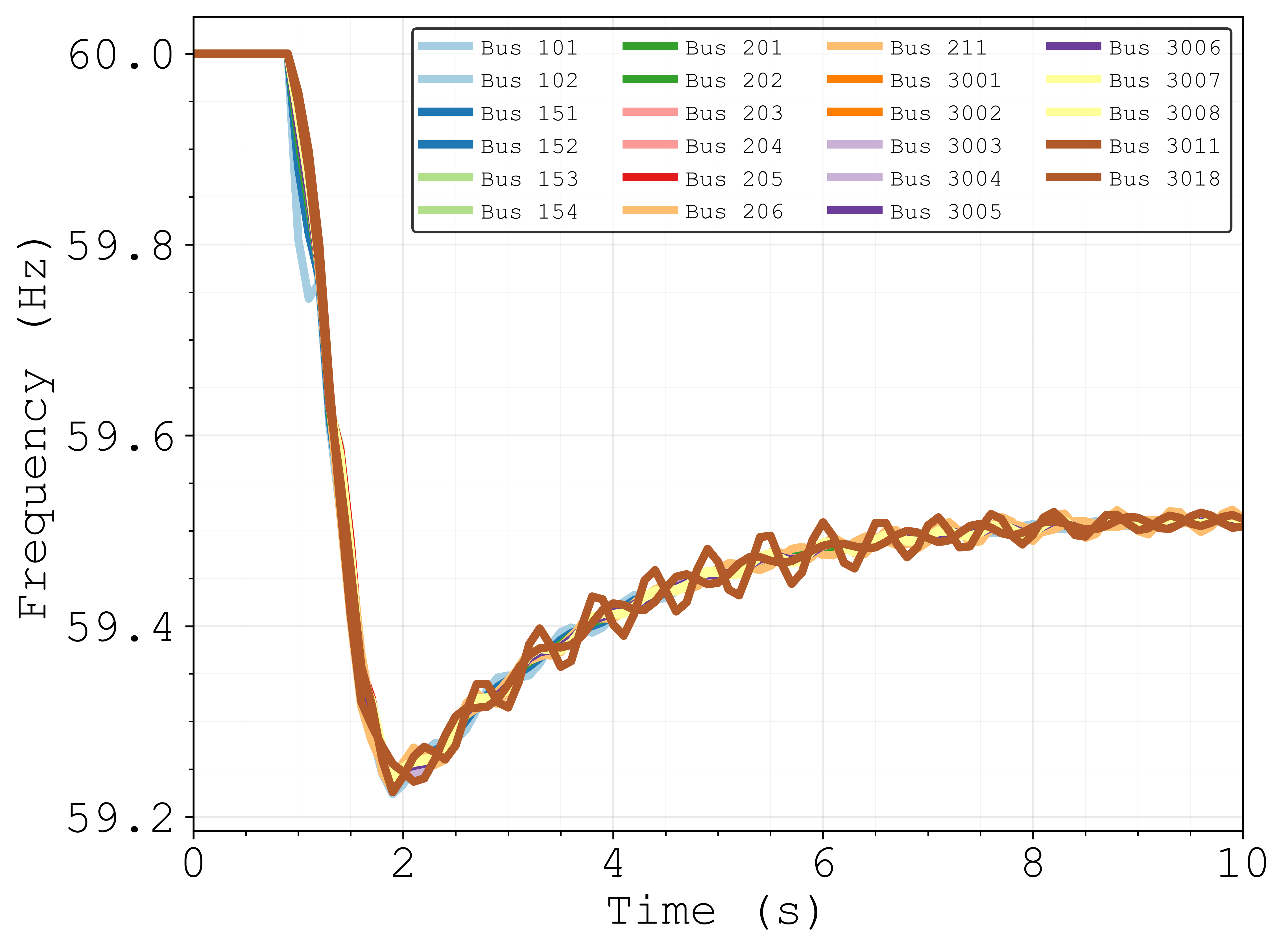}
    \caption{Frequency responses at all buses of the \texttt{savnw} system, showing that optimal load shedding satisfies the frequency bound with $\underline{\Delta \omega}$ set to 59.0~Hz.}
    \label{fig:4}
\end{figure}

In this section, we demonstrate how the \tos{} results translate into implementable UFLS settings. For discussion, we consider one representative case: the \texttt{savnw} network, where we impose a 25\% generation–load imbalance at $t=1 s$ by tripping the generators at buses~153 and~3018.
The trajectory-optimization algorithm identifies the load-shedding locations and stage activations. The resulting frequency trajectory stayed within the prescribed bounds, as shown in Fig.~\ref{fig:4}. We translate the optimal solution into implementable relay settings by post-processing the solved trajectory to extract frequency trigger points for each load and stage (Step~12 of the algorithm~\ref{alg:homotopy}). Table~\ref{tab:ufls_settings} reports the adaptive UFLS settings. At Stage~1, all candidate loads activate at thresholds between 59.45 and 59.49 Hz with a 20\% shedding amount. At Stage~2, loads 205, 3005, and 3007 trigger near 59.26–59.27 Hz with another 20\% shedding. The proposed settings halt the frequency decline without activating Stage~3, thereby preserving the more critical loads reserved for higher emergency conditions. These results confirm that the proposed workflow produces deployable UFLS settings that keep frequency within bounds while minimizing load shed, giving system operators a practical path to convert \tos{} outputs into adaptive UFLS settings.

\renewcommand{\arraystretch}{1.2}
\begin{table}[!t]
\centering
\caption{Adaptive UFLS settings for the \texttt{savnw} test system.}
\label{tab:ufls_settings}
\footnotesize
\setlength{\tabcolsep}{3pt}
\resizebox{\columnwidth}{!}{%
\begin{tabular}{lcccccc}
\toprule
\textbf{Load ID} &
\multicolumn{2}{c}{\textbf{Stage 1 Activation}} &
\multicolumn{2}{c}{\textbf{Stage 2 Activation}} &
\multicolumn{2}{c}{\textbf{Stage 3 Activation}} \\
\cmidrule(lr){2-3} \cmidrule(lr){4-5} \cmidrule(lr){6-7}
& Freq. (Hz) & Shed (\%) & Freq. (Hz) & Shed (\%) & Freq. (Hz) & Shed (\%) \\
\midrule
153  & 59.46 & 20 & --- & --- & --- & --- \\
154  & 59.49 & 20 & --- & --- & --- & --- \\
203  & 59.48 & 20 & --- & --- & --- & --- \\
205  & 59.48 & 20 & 59.26 & 20 & --- & --- \\
3005 & 59.45 & 20 & 59.27 & 20 & --- & --- \\
3007 & 59.46 & 20 & 59.26 & 20 & --- & --- \\
3008 & 59.47 & 20 & --- & --- & --- & --- \\
\bottomrule
\end{tabular}%
}
\end{table}

\subsection{Integer Feasibility within Trajectory Optimization}\label{subsec:results_2}
\begin{figure}
    \centering
\includegraphics[width=\columnwidth,keepaspectratio]{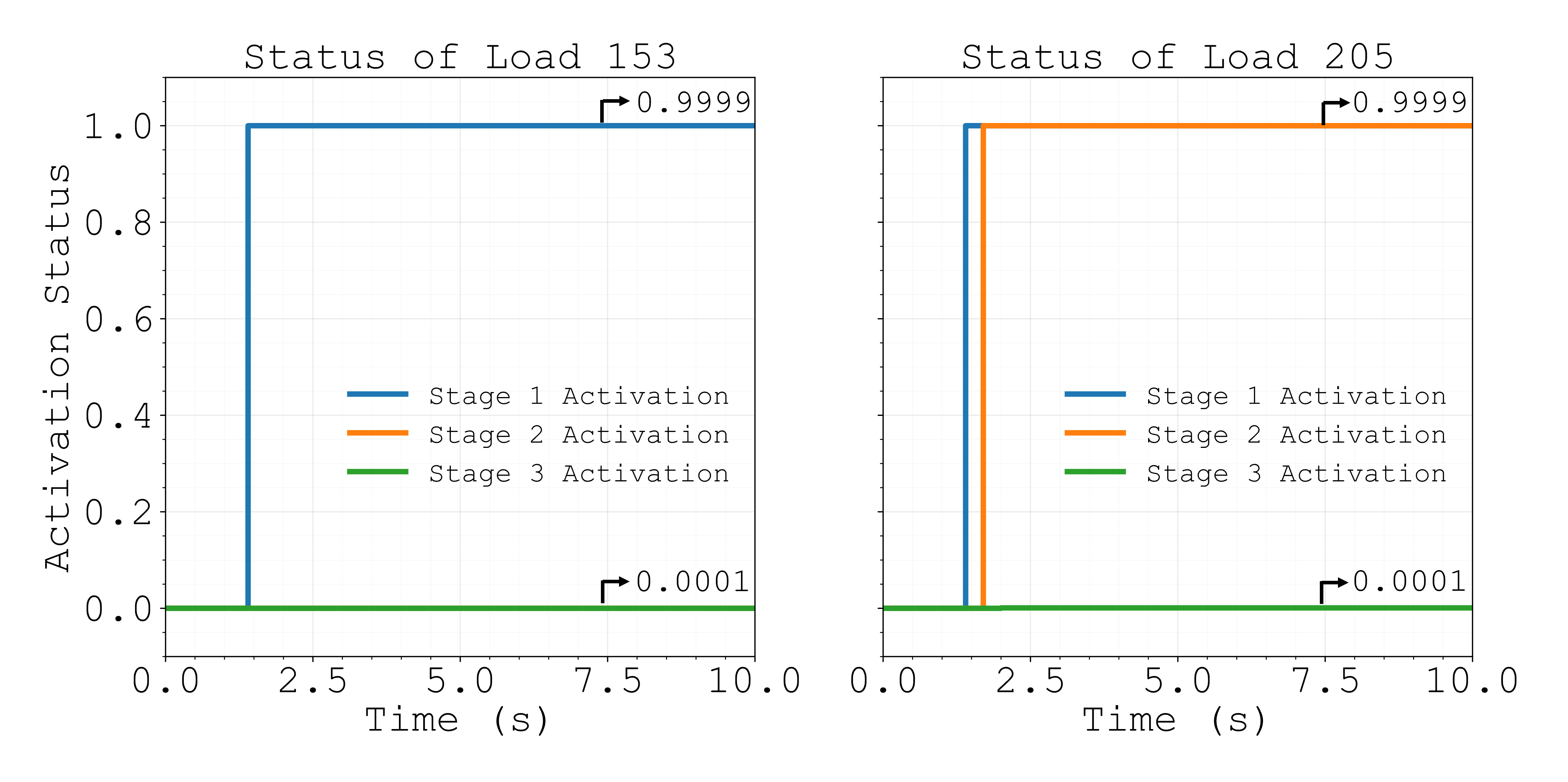}
    \caption{Near Integer feasible solution showing stage activations for loads at Bus 153 and Bus 205 from the \texttt{savnw} test case.}
    \label{fig:5}
\end{figure}

We present the switching results of the homotopy-embedded \tos{}, which achieves near integer-feasible solutions as shown in Fig.~\ref{fig:5}. To interpret the near-integer values, we assign a binary status of 0 or 1 when the relaxed variable lies within 0.0001\% of the respective binary limits, which provides a consistent mapping from continuous relaxation to discrete stage activations. These integer variables directly correspond to the activation status (on/off) of each load at each stage. Activating a stage executes the shedding of a predetermined block of load to maintain grid frequency within secure operational bounds.

We demonstrate the integer-feasible outcome of selected load relays in Fig.~\ref{fig:5}, which corresponds to the solution of the final subproblem for the \texttt{savnw} test case with bus frequencies shown in Fig.~\ref{fig:4}. For the load at Bus~153, the algorithm activates Stage~1 at 1.4 s, whereas for the load at Bus~10, it activates both Stage~1 at 1.4s and Stage~2 at 1.7 s to halt the frequency decline and support system recovery. We use this switch-status trajectory to compute the post-processed frequency thresholds reported in Table~\ref{tab:ufls_settings}.

\subsection{Performance Across Test Cases}
We evaluate the framework on multiple synthetic test cases under the largest credible generation loss to assess its scalability. Table~\ref{tab:test_cases_results} reports the nadir and settling frequencies along with the total number of continuous and switching decision variables. Iter. represents the total number of iterations accumulated across all epochs. Since each load is divided into three shedding stages, which in turn raises the number of variables that must attain integer feasibility within the optimization. These variable counts correspond to the chosen time discretization of the \tos{}. 

\begin{figure}
    \centering
\includegraphics[width=\columnwidth,keepaspectratio]{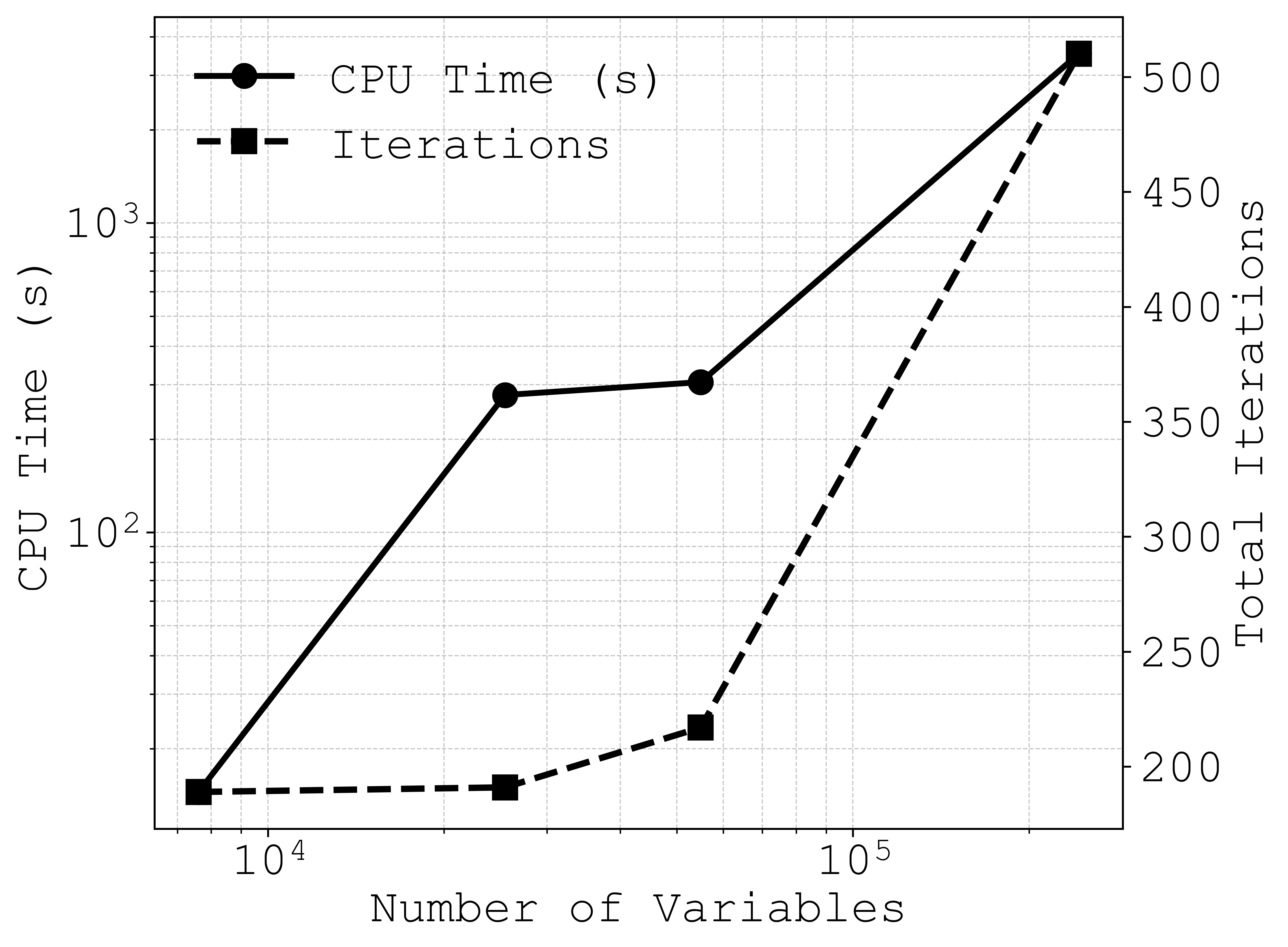}
    \caption{Solver runtime accumulated over all epochs and total iterations across test cases by accumulating continuous and switching variables.}
    \label{fig:6}
\end{figure}

We plot the total number of variables, including both continuous and switching variables, against the solver runtime and total iteration count in Fig.~\ref{fig:6}. The iteration count represents the total number of Ipopt iterations accumulated over all epochs. Time indicates the total runtime, including the time required for instantiating each model.
From Fig.~\ref{fig:6}, we observe that iteration count and CPU time increase nearly linearly with network size. All test cases can be executed with smaller time steps, but this would require significantly greater computational resources.
For each system, the adaptive UFLS scheme successfully halts the frequency decline and restores the system's frequency.  The results confirm that our framework consistently achieves feasible relay decisions while preserving system stability.
\renewcommand{\arraystretch}{1.4}
\begin{table}[!t]
\centering
\caption{Solver performance and frequency metrics for different dynamic test systems.}
\label{tab:test_cases_results}
\footnotesize
\setlength{\tabcolsep}{3.2pt}
\resizebox{\columnwidth}{!}{%
\begin{tabular}{lccccccc}
\toprule
\textbf{Test Case} &
\textbf{Nodes} &
\textbf{Nadir} & 
\textbf{Settling} &
\multicolumn{2}{c}{\textbf{\# of Vars}} & 
\textbf{Iter.} \\
\cmidrule(lr){4-5}
 & & (Hz) & (Hz) & $x$ & $z$ & & \\
\midrule
IEEE 59  & 59      & 59.43 & 59.58 & 6858  & 760  &   189 \\
ACTIVSg200     & 200 & 59.38 & 59.54 & 15832  & 9600 &  191  \\
ACTIVSg500     & 500 & 59.50 & 59.95 & 42508 & 12\,420 &  217  \\
Bench     & 1648 & 59.48 & 59.65 & 170045 & 73200 &  510  \\
\bottomrule
\end{tabular}%
}
\end{table}

\subsection{Penetration of IBR}
\begin{figure}
    \centering
\includegraphics[width=\columnwidth,keepaspectratio]{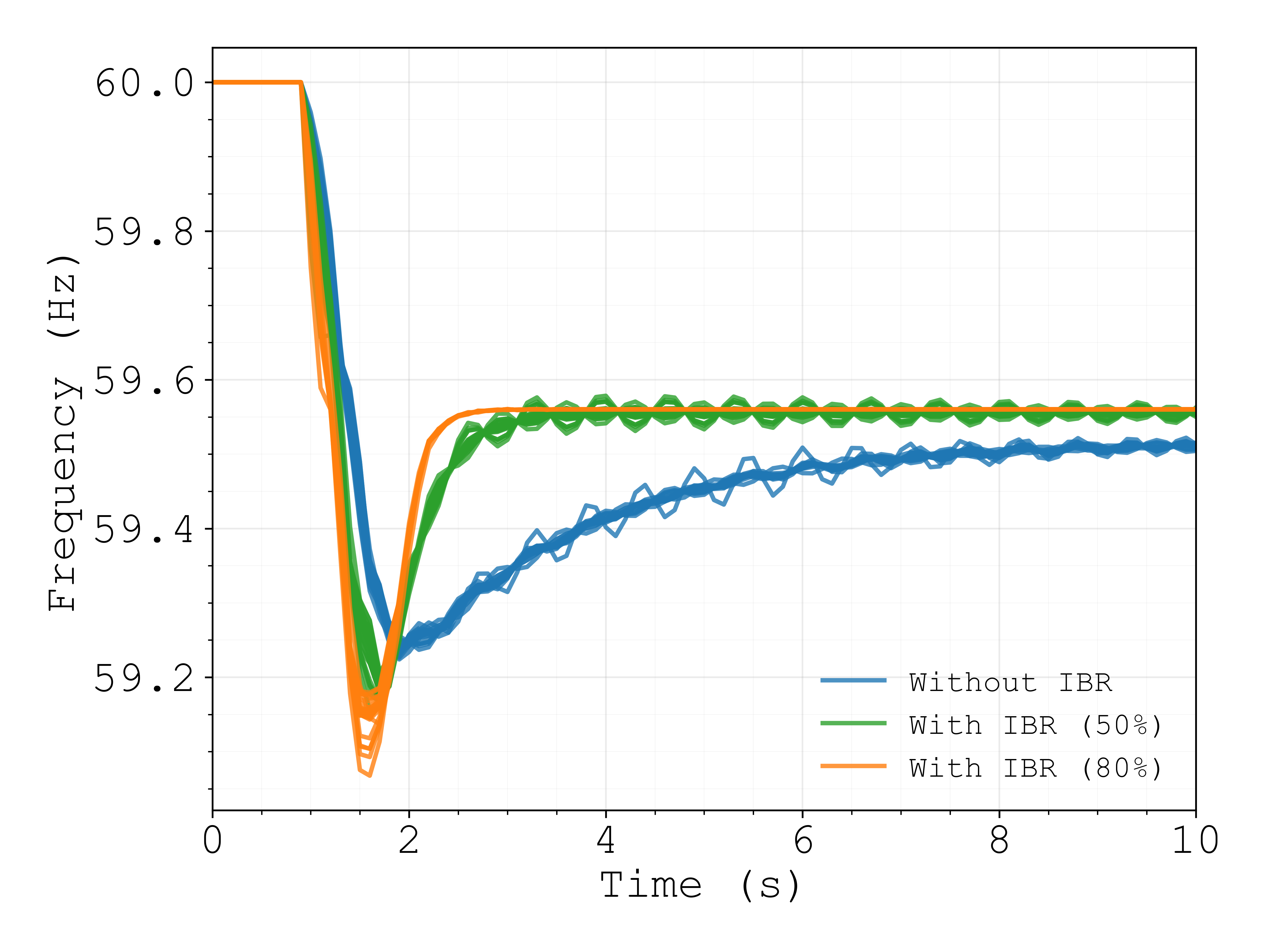}
    \caption{Frequency response and UFLS performance under increasing IBR penetration levels.}
    \label{fig:7}
\end{figure}

We evaluate the proposed adaptive UFLS optimization algorithm on the \texttt{savnw} test case, which includes six synchronous generators, under three scenarios: a baseline case with all synchronous generators retained; a 50\% inverter-penetration case in which we replace three synchronous generators with two GFM inverters and one GFL inverter; and an 80\% inverter-penetration case in which we replace five synchronous generators with two GFM inverters and three GFL inverters. High IBR penetration reduces the effective system inertia and causes the frequency to decline more rapidly following a disturbance. As shown by the frequency trajectories in Fig.~\ref{fig:7}, the 80\% IBR scenario reaches the nadir earlier and at a lower value than both the 50\% IBR scenario and the case without IBRs. In these simulations, the inverter response depends primarily on the selected droop gains and associated control time constants, which govern how quickly GFL and GFM inverters adjust active power during the transient. Despite the reduced time available to arrest the frequency fall, the optimization algorithm successfully coordinates the faster acting IBR  response with the load-shedding decision to stabilize the system.

Lastly, we modified the synthetic South Carolina 500-bus system to operate in a DER scenario by designating a subset of nodes as backfeeding buses. This configuration enables us to evaluate how the \tos{} coordinates load-shedding actions while accounting for DER backfeeding locations. Figure~\ref{fig:7} illustrates the spatial distribution of stage activations and DER nodes, demonstrating that the framework effectively arrests the frequency decline without selecting loads at DER-backfeeding buses.
\begin{figure}
    \centering
\includegraphics[width=\columnwidth,keepaspectratio]{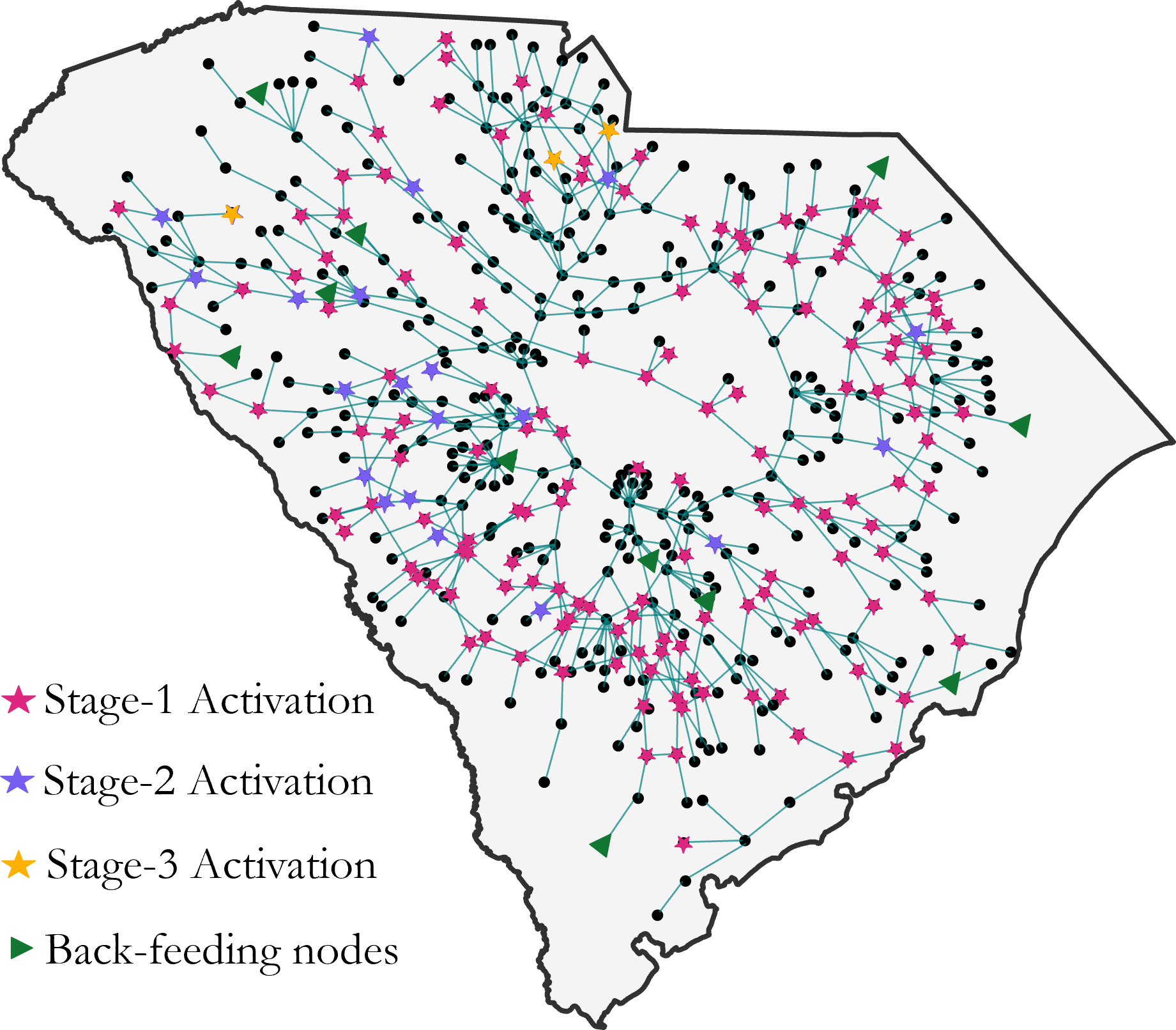}
    \caption{The synthetic South Carolina 500-bus system exhibits UFLS stage activations as certain buses operate as backfeeding nodes under high DER penetration.}
    \label{fig:8}
\end{figure}
\section{Conclusion \& Outlook}\label{sec:con}
We propose a novel adaptive UFLS scheme that overcomes limitations in existing methods and introduces the following key contributions:
\begin{enumerate}
    \item Embedding a full nonlinear AC dynamics within the trajectory optimization, avoiding linearization and accurately capturing transient frequency behavior.
    \item Enforcing binary switching decisions through a homotopy-driven method that achieves less than 0.0001\% deviation from ideal values and scales efficiently to networks exceeding 1500+ nodes with over 170k+ continuous and 73k+ switching variables.
    \item Developing a multi-stage formulation that coordinates sequential shedding actions and yields deployable relay setpoints for practical UFLS implementation.
\end{enumerate}

Overall, the proposed framework provides a promising solution for the adaptive UFLS design problem and can replace conventional UFLS schemes. The proposed framework is broadly applicable to other power system optimization and control problems that involve binary switching actions during transient events.
\section{Acknowledgment}\label{ack}
This research is funded by the National Science
Foundation through contract ECCS: 2330195 and Pacific Northwest National Laboratory DE-AC05-76RL01830. We acknowledge Mazen Elsaadany for contributions to power-system transient model development and Mads R. Almassalkhi for technical discussions.

\bibliographystyle{IEEEtran}
\bibliography{biblography}
\section{Appendix}\label{sec:appen}
We provide the parameters used in the model simulations in this appendix, while detailed network data for all test cases are available in the mentioned repository.
{\renewcommand{\arraystretch}{1.2}
\begin{table}[hbtp]
\centering
\footnotesize
\caption{List of parameters used in the numerical test.}
\label{tab:parameters}
\setlength{\tabcolsep}{5pt}
\begin{tabular}{lcl@{\hspace{12pt}}lcl}
\toprule
\textbf{Parameter} & \textbf{Value} & \textbf{Unit} &
\textbf{Parameter} & \textbf{Value} & \textbf{Unit} \\
\midrule
$f_{0}$                 & 60    & Hz   & $\alpha_{1}^{1}$        & 0.2   & --  \\
$a^{z}$                 & 0     & --   & $\alpha_{1}^{2}$        & 0.2   & --  \\
$a^{I}$                 & 1     & --   & $\alpha_{1}^{3}$        & 0.6   & --  \\
$a^{P}$                 & 0     & --   & $h$                     & 0.10  & s   \\
$b^{z}$                 & 1     & --   & $N$                     & 10    & --  \\
$b^{I}$                 & 0     & --   & $\tau^{s}$              & 0.30  & s   \\
$b^{P}$                 & 0     & --   & $\underline{\Delta\omega^{ss}}$ & 59.5 & Hz  \\
$T_{f}$                 & 0.02  & s    & $\overline{\Delta\omega^{ss}}$  & 60.5 & Hz  \\
$T_{w}$                 & 0.10  & s    & $\underline{\Delta\omega}$      & 59.0 & Hz  \\
\bottomrule
\end{tabular}
\end{table}
}

\end{document}